\documentclass{vldb}
\usepackage[T1]{fontenc}
\usepackage{balance}  
\usepackage{graphics} 
\usepackage{times}    
\usepackage{url}      
\usepackage{graphicx}
\usepackage{tabularx}
\usepackage{float}
\usepackage{color}
\usepackage{url}
\usepackage[noend]{algpseudocode}
\usepackage{algorithm}
\usepackage{verbatim}
\usepackage{mathtools}
\usepackage{caption}
\usepackage{subcaption}

\usepackage{amsthm}
\usepackage{thmtools}
\usepackage{xspace}
\usepackage{multirow}
\usepackage{mathrsfs}
\usepackage{mwe}
\usepackage{hyperref}
\usepackage{enumitem}
\usepackage[export]{adjustbox}
\usepackage[dvipsnames]{xcolor}

\hypersetup{colorlinks=true,allcolors=blue}

\newcommand{\name}{{PerfEnforce}\xspace}
\newcommand{\ie}{{\em i.e.}\xspace}
\newcommand{\eg}{{\em e.g.}\xspace}

\newcommand{\aka}{{\em a.k.a.}\xspace}

\newenvironment{packed_item}{
\begin{itemize}
   \setlength{\itemsep}{1pt}
   \setlength{\parskip}{0pt}
   \setlength{\parsep}{0pt}
}
{\end{itemize}}

\begin{document}

\title{\name: A Dynamic Scaling Engine for \\ Analytics with Performance Guarantees \vspace{-2ex}}

\numberofauthors{1}
\author{
\alignauthor
Jennifer Ortiz$^\dag$, Brendan Lee$^\dag$, Magdalena Balazinska$^\dag$,
and Joseph L. Hellerstein$^\ddag$ \\
         \affaddr{$^\dag$Department of Computer Science \& Engineering,$^\ddag$eScience Institute}\\
         \affaddr{University of Washington, Seattle, Washington, USA}\\
         \affaddr{\{jortiz16, lee33, magda\}@cs.washington.edu}, \affaddr{jlheller@uw.edu}\\
}
\maketitle

\begin{sloppypar}

\begin{abstract}
  In this paper, we present \name, a scaling engine designed
  to enable cloud providers to sell performance levels for data
  analytics cloud services. \name scales a cluster of virtual machines (VMs) allocated to a
  user in a way that minimizes cost while probabilistically meeting
  the query runtime guarantees offered by a
  service level agreement (SLA). With \name, we show \textit{how} to
  scale a cluster in a way that minimally disrupts a
  user's query session. We further show \textit{when} to scale the
  cluster using one of three methods: feedback control, reinforcement
  learning, or perceptron learning. We find that perceptron 
  learning outperforms the other two methods when
  making cluster scaling decisions.
\end{abstract}

\fontsize{10pt}{12pt}\selectfont
\textheight 9in
\oddsidemargin 0in
\topmargin 0in
\headheight 0in
\headsep 0in

\vspace{-0.25cm}
\section{Introduction}

A variety of systems for data analytics are available as cloud services today, including Amazon Elastic MapReduce (EMR), Amazon Redshift~\cite{amazonaws}, Azure's HDInsight~\cite{azure}, and several others. While these services greatly facilitate access to compute
resources and data analytics software, they remain difficult for users to tune in terms of cost and performance. Users choose a price-performance trade-off by selecting a desired number and type of service instances.  It is well-known, however, that users have difficulty determining their resource needs and often attempt many configurations before finding a suitable one~\cite{Herodotou:11a}. Some systems do not offer any configuration choices. Google BigQuery~\cite{bigquery} is one example. These systems, however, deprive users of the ability to adjust how much money they want to spend on an analysis at the expense of some loss in performance. There exist systems that can help select a cluster configuration~\cite{Herodotou:11a, Jalaparti:12}.  However, these prior methods are specific to MapReduce engines and also require profile runs of each job. In contrast, we target exploratory analytics, where users interactively submit ad-hoc queries and we develop an approach that can easily be applied to any big data system.



\begin{figure}[t]
  \centering
  \includegraphics[width=\linewidth]{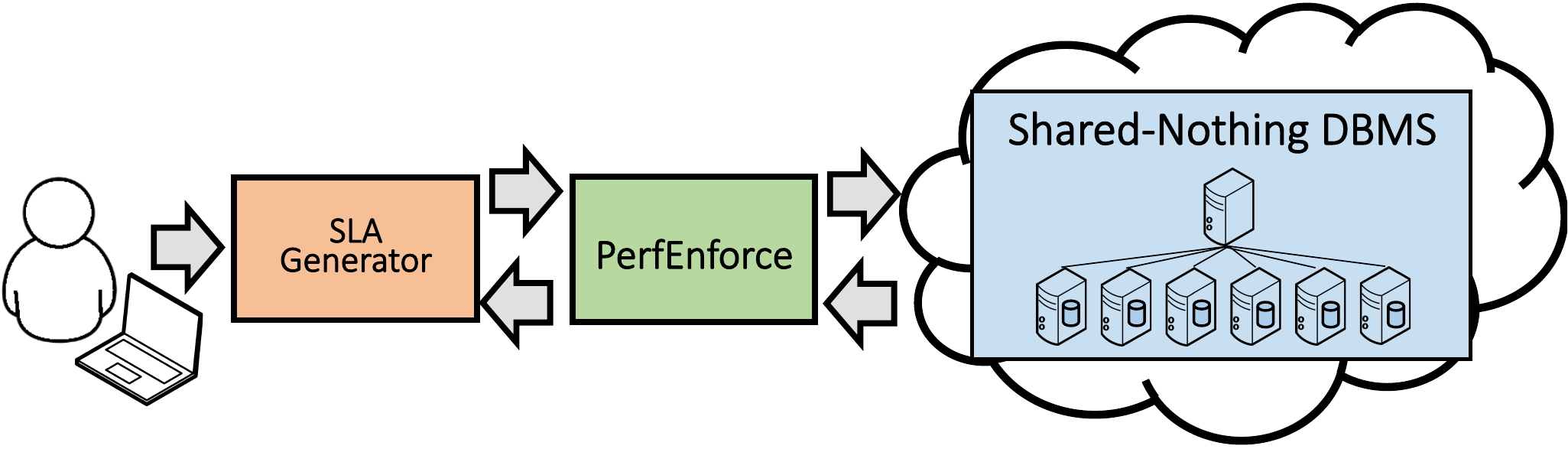}
  \caption{\name deployment: \name sits on top of an elastically scalable big data management system  in support of performance-oriented SLAs for cloud data analytics provided by an SLA Generator.}
  \label{fig:model}
  \vspace{-0.8cm}
\end{figure}

Performance-centric service level agreements (SLAs)~\cite{Papaemmanouil:12,ortiz:15} have  been proposed in
response to the above limitations. With this approach, a user buys a given \textit{performance level} (query latency) rather than an amount of resources. However, a fundamental challenge with performance-centric SLAs, is how to guarantee the performance that the user purchases.  This problem is important because, for performance-based SLAs to be meaningful, they must come with concrete performance guarantees. For example, the SLA could specify that 90 percent of the user queries will execute within their posted runtime. If the SLA is violated, the user receives a predefined compensation.

In this paper, we develop a system called \name that works with a cloud service to meet the goals of a performance-based SLA.  \name is designed for data management systems that support data analytic workloads (\eg, Myria~\cite{halperin:14}, Spark~\cite{spark}, Impala~\cite{Liu:03}, EMR~\cite{amazonaws}). Additionally, \name targets cloud services that follow a model such as that of Amazon EC2~\cite{amazonec2} and Azure HDInsight~\cite{azure}, where each user performs her analysis using a separate set of virtual machines (VMs). In this paper, we do not address the problem of how to generate a performance SLA, which was the focus of prior work including our own~\cite{ortiz:15}. PerfEnforce assumes that the SLA exists and takes as input a set of pairs: ${ (q_i, t_i) }$, where $q_i$ is a query submitted by the user and $t_i$ is the SLA runtime associated with that specific query. 

\autoref{fig:model} shows the system architecture: The user first purchases a performance SLA generated by an SLA Generator. PerfEnforce provisions the cluster of VMs on behalf of the user by ingesting the user's data into the cluster and monitoring the execution of the user's queries. To guarantee the query runtimes associated with an SLA, \name resizes the cluster in between queries either in a proactive or reactive approach. With a proactive approach, PerfEnforce decides to scale based on how well it met previous SLA deadlines. In a reactive approach, PerfEnforce decides whether to rescale the cluster before executing each incoming query. \name's goal is to select the cheapest configuration possible in order to meet the SLA runtimes.


During the user's query session, PerfEnforce faces two key technical challenges: \textit{how} to rescale the cluster and \textit{when} to rescale it.

Quickly scaling a cluster (either up or down) to meet SLA guarantees or save costs is not trivial.  Re-allocating resources during data analysis can be disruptive to the analysis if it requires significant data shuffling. At the same time, data replication in preparation for quick scaling can increase setup costs, which are known to be highly undesirable~\cite{Gupta:15}. Deployments that separate between compute and data nodes to accelerate setup and cluster configuration changes can either negatively impact query runtimes or significantly increase costs.  In this paper, we empirically evaluate a set of elastic scaling methods and compare them in terms of initial setup time, the storage type, time to change the cluster configuration between queries, query execution time, and total cost. We demonstrate the above challenges associated with inexpensive and rapid scaling and show that careful data placement and partial replication offer a practical solution to the problem.

The second challenge is when to decide to scale the cluster up or down.  Several systems have recently studied performance guarantees through dynamic resource allocation in storage systems~\cite{Lim:10} using feedback control, or in transaction processing systems~\cite{Konstantinou:12} using reinforcement learning. In this paper, we show how to apply feedback control and reinforcement learning to the problem of query time guarantees for data analytics. We experimentally demonstrate, however, that these approaches do not work well in this context because query time estimation errors can vary significantly for consecutive queries and errors can be in either direction (under- or over-estimation of query times). As a result, during a single user session the system does not converge to a single cluster size but
instead needs to make resource allocation decisions separately for each query. Based on this observation, we develop a third cluster-scaling algorithm. Our approach uses perceptron learning: As the user executes queries, PerfEnforce continuously updates its model of query time estimates. Perceptron learning has the double benefit of quickly adapting to the user's recent query workload and current system conditions. In addition, we apply this approach without having to build an analytical model of the underlying system. Features of the system are simply fed into the model and query latencies are adaptively learned. PerfEnforce then uses this model to select the most appropriate cluster size separately for each query. We show experimentally that this approach delivers better quality of service and is more cost-effective than either feedback control or reinforcement learning.

In summary, we make the following contributions:
\begin{packed_item}

\item We develop PerfEnforce, a dynamic scaling engine for data analytics services (\autoref{sec:overview}).

\item We quantitatively evaluate different data placement and cluster re-sizing methods  (\autoref{sec:scaling_methods}).


\item We adapt well-known resource scaling algorithms based on feedback control and reinforcement learning to the problem of query
time guarantees for data analytics (\autoref{sec:scaling:reactive}).

\item We develop a new resource scaling algorithm based on perceptron learning (\autoref{sec:scaling:proactive}).

\item We study the performance of the three scaling algorithms through experiments with the Myria~\cite{halperin:14} shared-nothing DBMS and the Amazon EC2 cloud~\cite{amazonec2} (\autoref{sec:scaling_algorithms_eval}). 

\end{packed_item}

\vspace{-0.25cm}
\section{Related Work}

\textbf{Performance Guarantees in Data Analytics} Performance guarantees have traditionally been the focus of real-time database systems~\cite{Kao:95}, where the goal is to schedule queries in a fixed-size cluster to ensure they meet their deadlines. More recently, 
dynamic provisioning and admission control methods have enabled OLAP and OLTP systems to make profitable choices with respect
to performance guarantees~\cite{Chi:11a, Chi:11b,Xiong:11a}, possibly postponing or even simply rejecting queries. \name's goal instead is to scale the cluster with minimal delay to meet SLA guarantees.



\textbf{Multi-Tenant Performance Guarantees} An active area of research in multi-tenant cloud DBMS systems is \textit{tenant packing} ~\cite{Elmore:13, Mahmoud:13, Liu:13}, or how best to colocate tenants on a shared set of machines or even DBMS instances. In contrast, we focus on the independent database user who spins up his own private cluster in the cloud. We seek to  minimize the
size of that cluster while meeting SLA runtime guarantees.



\textbf{Query Runtime Prediction} Previous work has relied on classification and regression techniques to determine whether a query will miss or meet a deadline~\cite{Xiong:11a}, building gray-box performance models~\cite{Gandhi:16}, using historical traces of previous workloads~\cite{Ferguson:12} or running smaller samples of the workload with a low overhead~\cite{Venkataraman:16}. Most closely
related is work by Herodotou et. al. ~\cite{Herodotou:11a}, which assumes a previously profiled workload from the user in order to predict the runtime of that program against different sized clusters. Work by Jalaparti et. al.~\cite{Jalaparti:12} focuses on generating resource combinations given performance goals from the user. Instead of building a white-box or analytical model, we focus on using a model that does not require an extensive understanding of a single system. We also focus on interactive, ad-hoc queries for which there are no prior profiles.

\textbf{Elasticity} Cloud providers offer the ability to scale a database application~\cite{amazonaws,azure}. However, they require users to manually specify scaling conditions through vendor-specific APIs. This requires expertise and imposes the risk of resource over-provisioning. Moreover, these scaling features can be costly, as some of these actions are subject to service downtimes and may take several minutes to complete (such as data rebalancing)~\cite{amazonaws}.

Most academic work on elastic systems focuses on OLTP workloads~\cite{Das:11, Stonebraker:13, Vo:10} and thus develops new techniques for tenant database migration~\cite{Elmore:14}, data re-partitioning while maintaining consistency~\cite{Minhas:12} or automated replication~\cite{Vo:10}. In these systems, the goal is to maximize aggregate system performance, while our focus is on a per-query performance guarantees. 


\vspace{-0.25cm}
\section{PerfEnforce Overview}
\label{sec:overview}

In this section, we present an overview of \name: How \name interacts with the other components of a cloud service, what it assumes about the cloud service, what it takes as input, and its internal optimization goal.

\subsection{\name API} \label{sec:api}

\name is designed to work with a DBMS for data analytics, an SLA Generator, and a cloud service.  \autoref{fig:model} illustrates how \name interacts with these components. When a user begins her query session, she first purchases a \textit{performance-level} given by the SLA Generator. Given the performance-level selected, the SLA Generator provides an initial cluster size to \name, $init_c$, to begin the session. \name then monitors the query session to rescale if necessary.

\begin{table}[t]
\caption{\name's API.}
\label{tab:api}
\begin{tabular}{l|l}
\textbf{Function name and parameters} & \textbf{Returned value}\\ 
\hline
\textbf{Initialize} ($D$, $init_c$, $configs$) & $id$ \\
\textbf{Query} ($id$, $q$, $t_{sla}$)  & void \\
\textbf{Terminate} ($id$)  & void \\
\end{tabular}
\end{table}

\name exposes an API with three methods as shown in \autoref{tab:api}. The SLA Generator calls these methods. The \texttt{Initialize} method takes as input the user's data $D$, an initial cluster size $init_c$, and also the set of cluster sizes, $configs$. This method deploys an initial set of virtual machines (VMs), starts the DBMS, and ingests the data, $D$.   The method returns a unique session identifier, \texttt{id}. Subsequently, each call to the method \texttt{Query} passes the SQL query, $q$, to execute in the session $id$ and the SLA time, $t_{sla}$ associated with this query. The \texttt{Terminate} call deletes a previously deployed cluster. As \name scales the cluster during the query session, it keeps its size within the minimum and maximum values specified in the set $configs$.


In addition, \name requires the following functionality from the underlying DBMS system: (1) Ability to add and remove workers dynamically and (2) control over the way the data is organized in the cluster. \name can still work with a system that does not have complete control of the data layout, but this may impact performance as we explore in \autoref{sec:scaling_methods}.


\subsection{\name's Optimization} \label{sec:problem}

Given a query session \textit{Q}, with queries \textit{$q_0$} through \textit{$q_n$} and a set of cluster sizes \textit{configs}, \name optimizes what we call the \textit{Performance Ratio (PR)} of a query session. We define $PR$ as: 

\begin{equation} \label{eq:PR}
 \mathrm{PR}(Q) = \frac{1}{n} \sum_{q=0}^n \frac{t_{real}(q_i)}{t_{sla}(q_i)}
\end{equation}

In \autoref{eq:PR}, \textit{$t_{sla}(q_i)$} and \textit{$t_{real}(q_i)$} represent the SLA and actual runtimes of a query $q_i$, respectively. In order to neither waste cluster resources nor violate SLA runtimes, \name's goal is to maintain $PR(Q)$ as close to 1 as possible.



In \autoref{sec:scaling_algorithms}, we show how different cluster-scaling algorithms, yield different $\frac{t_{real}(q_i)}{t_{sla}(q_i)}$ distributions (See \autoref{fig:dist}). The best cluster scaling algorithm is one that (1) yields a tight distribution close to 1.0, which ensures that most query runtimes stay close to the promised ones from the SLA and (2) achieves this goal at a low service cost.  We define the Cost of Service (CS) as:

\begin{equation} \label{eq:CS}
\mathrm{CS}(Q) = \sum_{i=0}^n cost(q_i)
\end{equation}

For \autoref{eq:CS}, $cost(q_i)$ is defined as the cost of virtual machines used to execute $q_i$.

\vspace{-0.25cm}
\section{Data Organization} \label{sec:scaling_methods}

In this section, we define and evaluate how to store data on disk to (1) ingest data quickly in preparation for the query session and (2) facilitate scaling with minimal interruptions during the query session. 
  PerfEnforce targets cloud services that execute the data management and analytics software in a separate set of VMs for each tenant.  In that context, data can be stored in the local storage of each VM or in a separate storage system available over the network.




When a user starts her query session, PerfEnforce prepares an initial set of $init_c$ VMs. Additionally, PerfEnforce prepares the system to resize itself to any cluster size in the set given by $configs$.


PerfEnforce has many choices to organize the user's data and scale resources up and down. First, we introduce the available storage types in Amazon AWS and evaluate them on data ingest and data read times. We then present different data placement methods and evaluate them based on latency to first query and disruption due to cluster resizing.

\begin{figure*}[t]
  \begin{subfigure}{.33\textwidth}
  \centering
  \includegraphics[width=\linewidth]{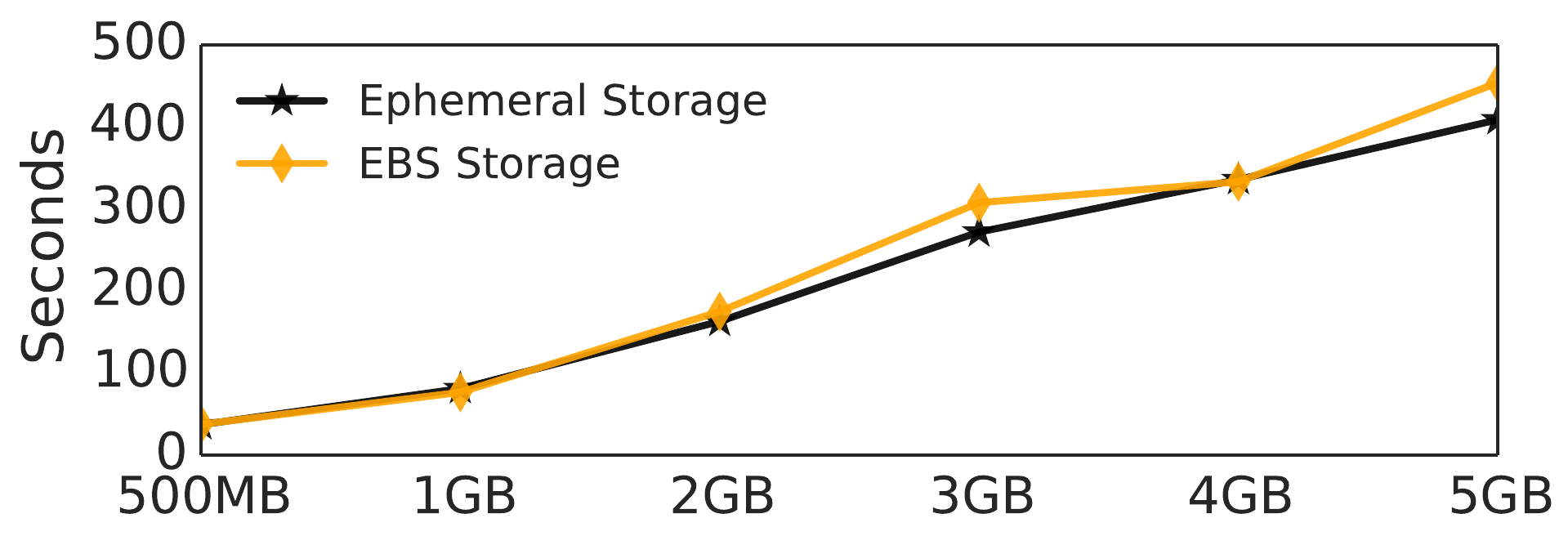}
  \caption{Time to Ingest Data}
  \label{fig:data_ingest_storage}
  \end{subfigure}
  \begin{subfigure}{.33\textwidth}
  \centering
  \includegraphics[width=\linewidth]{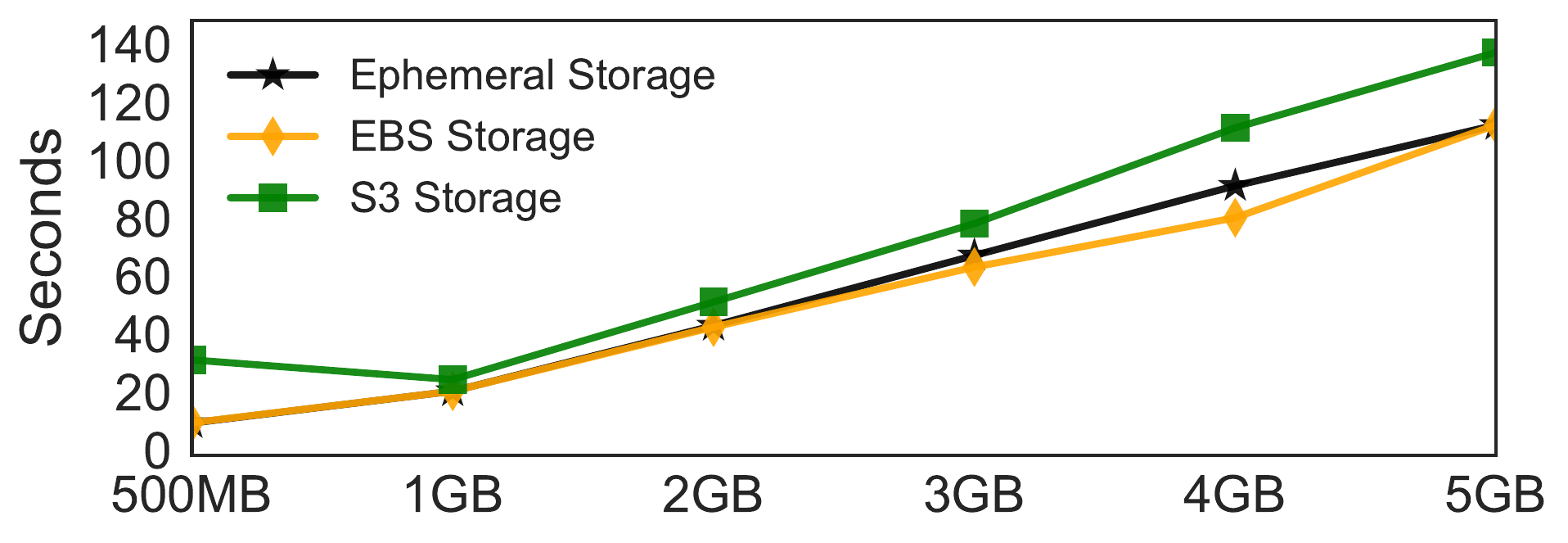}
  \caption{Time to Read a Table}
  \label{fig:time_read_table}
  \end{subfigure}
  \begin{subfigure}{.33\textwidth}
  \centering
  \includegraphics[width=\linewidth]{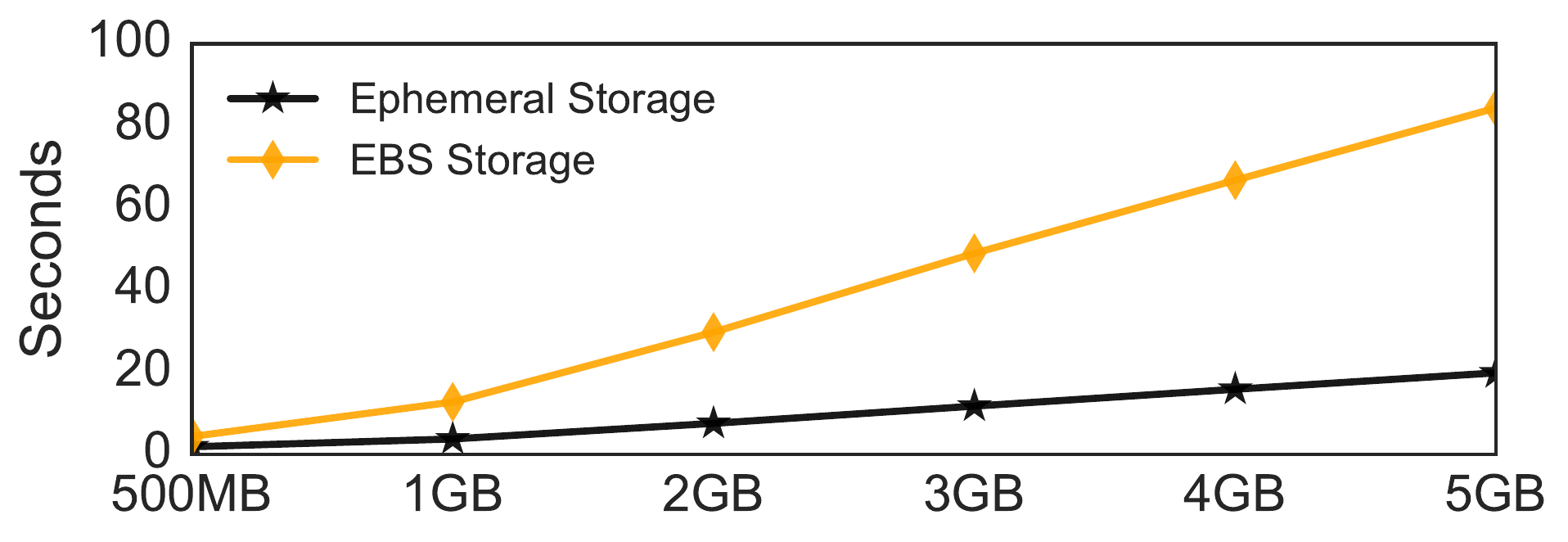}
  \caption{Time to Read a Column}
  \label{fig:time_read_column}
  \end{subfigure}  
  \caption{Evaluating Storage Options}
  \vspace{-0.5cm}
\end{figure*}

\subsection{Local, Networked, and Shared Storage}

A big data system can read and write data from a variety of sources once a set of VMs is provisioned.  In the concrete case of the Amazon cloud, these include, but are not limited to, Amazon Simple Storage Service (S3), Amazon EC2 Instance Store (Ephemeral) and Amazon Elastic Block Storage (EBS)~\cite{amazonec2}. To evaluate these storage options, we provision one m3.large node (4 ECU, 7.5 GB Memory) and read and write the lineorder table from the TPC-H Star Schema Benchmark (SSB)~\cite{oneil:09} dataset. For EBS, we use a general purpose SSD (gp2) type. For S3, we make sure that the S3 bucket and VM are within the same region.  \autoref{fig:data_ingest_storage} shows the time to ingest data. There is no ingest time for S3 as nodes can read data directly from that storage service during query execution. Ingest times are nearly identical for the other two storage systems. \autoref{fig:time_read_table} and \autoref{fig:time_read_column} show the time to read data.  While reading an entire table takes the same amount of time between ephemeral and EBS, reading a subset, such as one column, is significantly faster when using ephemeral storage compared to EBS. Reading tables from S3 takes slightly longer and it is not possible to read a single data column from S3.  Kossmann et. al. \cite{Kossmann:10} also report, though in the context of OTLP workloads, that EBS and ephemeral storage achieve similar performance. Compared to S3, both ephemeral and EBS have the additional advantage of caching data locally during the query session and performing local joins without having to reshuffle data when tables are partitioned on their join attribute. To minimize $CS(Q)$ and maximize query performance, we opt to use ephemeral storage since its price is included in the VM price and its performance is highest when reading subsets of the data. In the remainder of the paper, we use only ephemeral storage. We show next how to ensure that data ingest and cluster reconfiguration times both remain low with this storage option.

\subsection{Data Placement Strategies}

PerfEnforce replicates small dimension tables across all workers (\aka nodes) while partitioning large fact tables. Small tables take a negligible amount of time to copy over to a new worker. As such, any approach for cluster scaling works with small tables. The question is how to best manage cluster scaling for large tables.

Workers responsible for reading data constitute the \textit{data storage layer} of the system. The \textit{compute layer} are the workers that execute query operators such as joins or aggregates. We first consider the case where each worker serves as both a data and a compute node: \ie, when running queries with N workers, each worker stores and processes $\frac{1}{N}$th of the data. 



\textbf{Shuffled-Scaling} In this method, each large table is first uniformly partitioned across the initial set,  $init_c$, of workers using hash-, range-, or random data partitioning. To resize the cluster to a different configuration $c'$, PerfEnforce issues a query that reads the table, shuffles it, and re-materializes it across the updated set of $c'$ workers.
An important optimization is for workers to reshuffle only the minimal amount of data needed to rescale. This can be done by using consistent hashing~\cite{Karger:97} or simply using mini partitions as follows: Let $P_R = \{p_{r_0}, p_{r_1}, ..., p_{r_n}\}$ represent the partitions of relation, $R$, where $n$ is the number of nodes in configuration $init_c$. Each partition, $p_{r_i}$ is assigned to one node from configuration $init_c$ and is further split into $j$ mini partitions. In order to scale from $init_c$ to $c'$, each node needs to only read and shuffle a fraction of its mini partitions. For
example, when resizing from 2 to 4 workers, each of the original two workers must reshuffle half of its mini partitions across the two new workers.






\textbf{Static-Replicated} To avoid data re-shuffling upon cluster rescaling, \name can ingest multiple copies of each big table. Each copy is uniformly partitioned across a subset of machines that corresponds to one configuration in $configs$. For example, one copy of a table is partitioned across four workers, a second copy is partitioned across six workers, a third across eight, etc.

As an optimization, instead of ingesting multiple full copies of each big table in sequence for each configuration in $configs$, \name can, once again, use either mini-partitions or consistent hashing to only replicate a minimum amount of data.
For example, assume $configs = \{2, 4\}$. \name first partitions relation R across four workers as $P_R = \{p_{r_1}, p_{r_2}, p_{r_3}, p_{r_4}\}$. To generate a 2-worker partition, $P_{R'}$, \name copies $p_{r_3}$ and  $p_{r_4}$ onto workers $r_1$ and $r_2$ respectively. We call this approach \textit{Static-Replicated Chunks}.

\textbf{Dynamic-Scaling} The final approach distinguishes between sets of compute nodes, $C_{compute}$, and data nodes, $D_{data}$. \name uniformly ingests the tables into the number of assigned data nodes, $D_{data}$. The data layer remains fixed and never changes in size. Instead of re-materializing a table for a new configuration $c'$, \name only reads data from $D_{data}$, and shuffles the data to the $C_{compute}$ nodes (s.t. $|C_{compute}| = c'$ ) in order to finish the computation of the query. We consider two cases of Dynamic-Scaling: $|D_{data}| < |C_{compute}|$ and $|D_{data}| > |C_{compute}|$, which we call \textit{Dynamic-Small} and \textit{Dynamic-Large} respectively. For example, when provisioning a Dynamic-Small cluster, the system can fix the number of data nodes to 4 workers and only scale the number of compute nodes to range from 5 to 10 nodes. This can be advantageous if the user workload is CPU-bound. In a Dynamic-Large cluster, the system spreads the data thinly to many data nodes, which would then shuffle data to a smaller number of compute nodes to finish the query computation. Keeping the data thin is beneficial particularly for IO-bound workloads. 

Among the three techniques above, Shuffled-Scaling risks imposing high overheads when changing between cluster configurations. Static-Replicated scaling risks slowing down the initial data ingest time. Dynamic-Scaling is more costly (as one has to pay for both data and compute nodes).  We evaluate these techniques next.

\vspace{-0.25cm}
\subsection{Data Placement Evaluation} \label{sec:scaling_methods_eval}

We run \name on an Amazon EC2 cluster. Each node is an m3.large (4 ECU, 7.5 GB Memory) type. We consider five types of possible configurations, $configs=\{4,6,8,10,12\}$. For our underlying database management system, we use Myria~\cite{halperin:14} as it provides the ability to easily control data placement. Myria uses PostgreSQL as its node-local storage subsystem.

For our dataset, we use the TPC-H Star Schema Benchmark (SSB)~\cite{oneil:09}. This dataset consists of one fact table (lineorders) and a set of four smaller dimension tables. In total, the dataset is approximately 10GB, containing 5 tables and 58 attributes. We choose this dataset size because multiple Hadoop measurement papers report 10GB as a median input dataset analyzed by users~\cite{Ren:13}. For our query pool, we generate a set of approximately 900 select-project-join queries using our open-source PSLAManager tool~\cite{ortiz:15}. 

\begin{figure}[t]
  \centering
  \includegraphics[width=.8\linewidth]{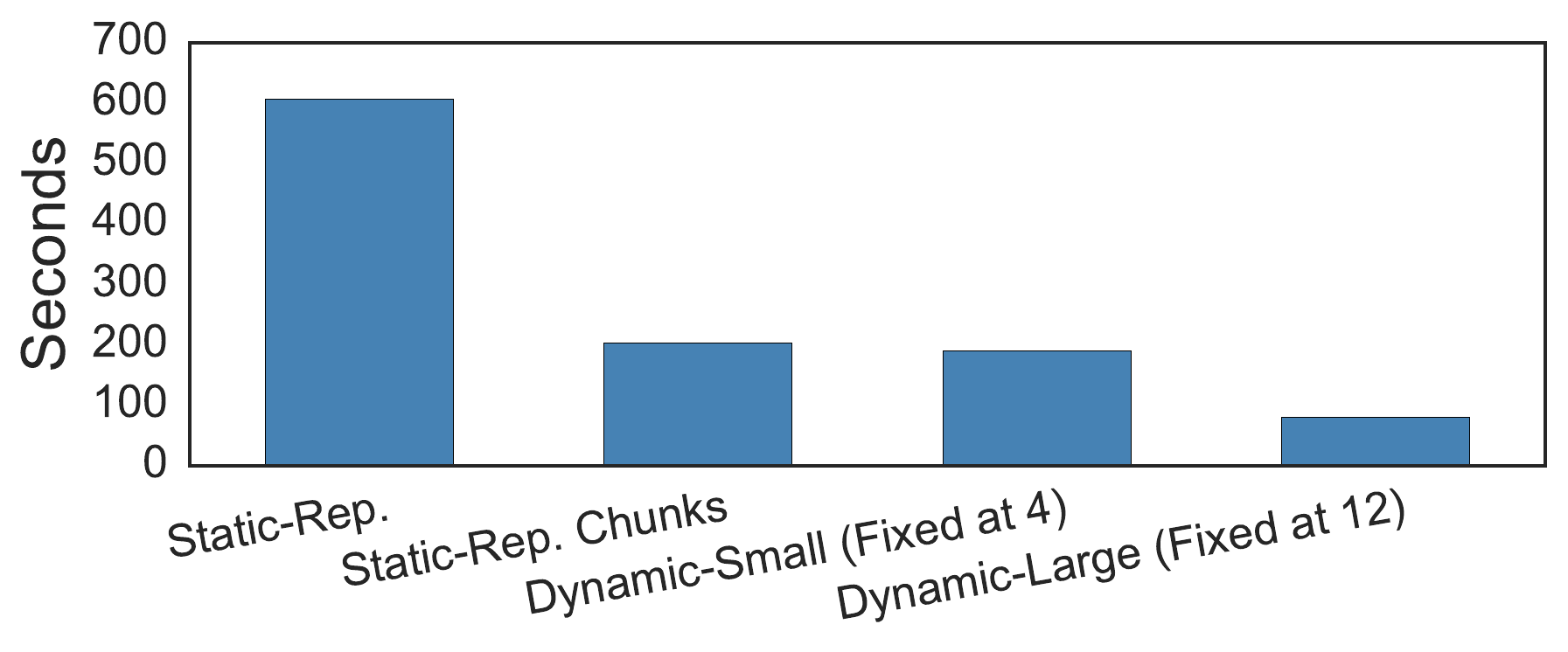}
  \caption{Time to Ingest Data for Scaling Methods}
  \label{fig:data_ingest_runtimes}
  \vspace{-0.5cm}

\end{figure}

\textbf{Data Ingest Runtime}
Given that \name operates as a cloud service, it must prepare and ingest the data efficiently in order to allow the user to begin the query session quickly. We consider the time it takes for each scaling method to ingest the TPC-H SSB dataset. In \autoref{fig:data_ingest_runtimes}, we display the runtimes for ingesting data using Static-Replicated, Static-Replicated Chunks, Dynamic-Small or Dynamic-Large methods. 

Ingesting for Static-Replicated takes approximately 606 seconds. This method takes the longest as it requires five copies of the lineorder table. For Dynamic-Scaling, we show the ingest runtimes for 4 and 12 fixed nodes. Ingesting data for the Static-Replicated Chunks method is comparable to ingesting data for the smallest configuration in $configs$ (4 workers). The Shuffled-Scaling method (not shown) takes the same time as either Dynamic-Small or Dynamic-Large, depending on the number of nodes in $init_c$.


In general, the bottleneck for ingest time largely depends on either the fixed number of data nodes selected for Dynamic-Scaling or the smallest configuration size that exists in $configs$ for Static-Replicated Chunks. Most importantly, the latter method provides the benefit of a replicated set of tables without the data ingest overhead associated with full data replication.

\textbf{Delay When Changing Between Configurations} Another factor to consider is the time it takes to switch between configurations in $configs$. For Static-Replicated and Static-Replicated Chunks, the multiple copies of the data allow for immediate scaling. For Dynamic-Scaling, the only factor that needs to change are the number of compute nodes. For both of these scaling methods, there is no delay when scaling the system, as no data materialization is required when switching between configurations.

\begin{figure}[t]
  \begin{subfigure}{.5\textwidth}
  \centering
  \includegraphics[width=.8\linewidth]{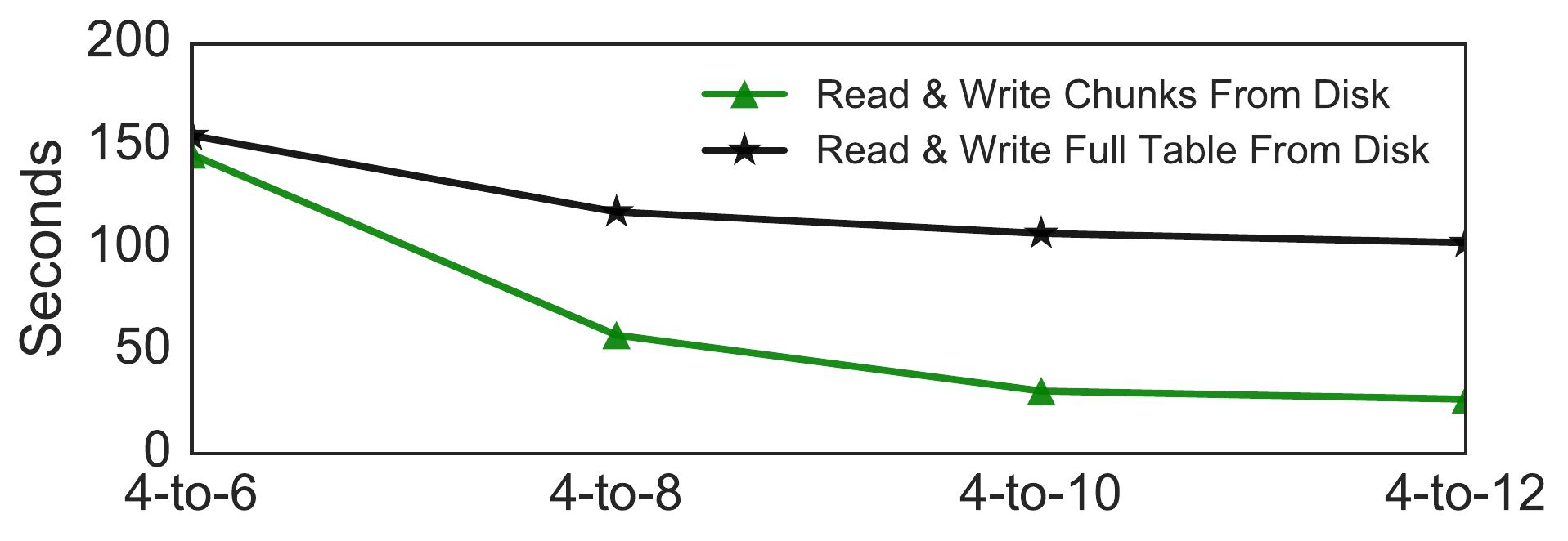}
  
  \caption{Time to Change Configurations from 4 Workers}
  \label{fig:reconfiguration_4Workers}
  \end{subfigure}

  \begin{subfigure}{.5\textwidth}
  \centering
  \includegraphics[width=.8\linewidth]{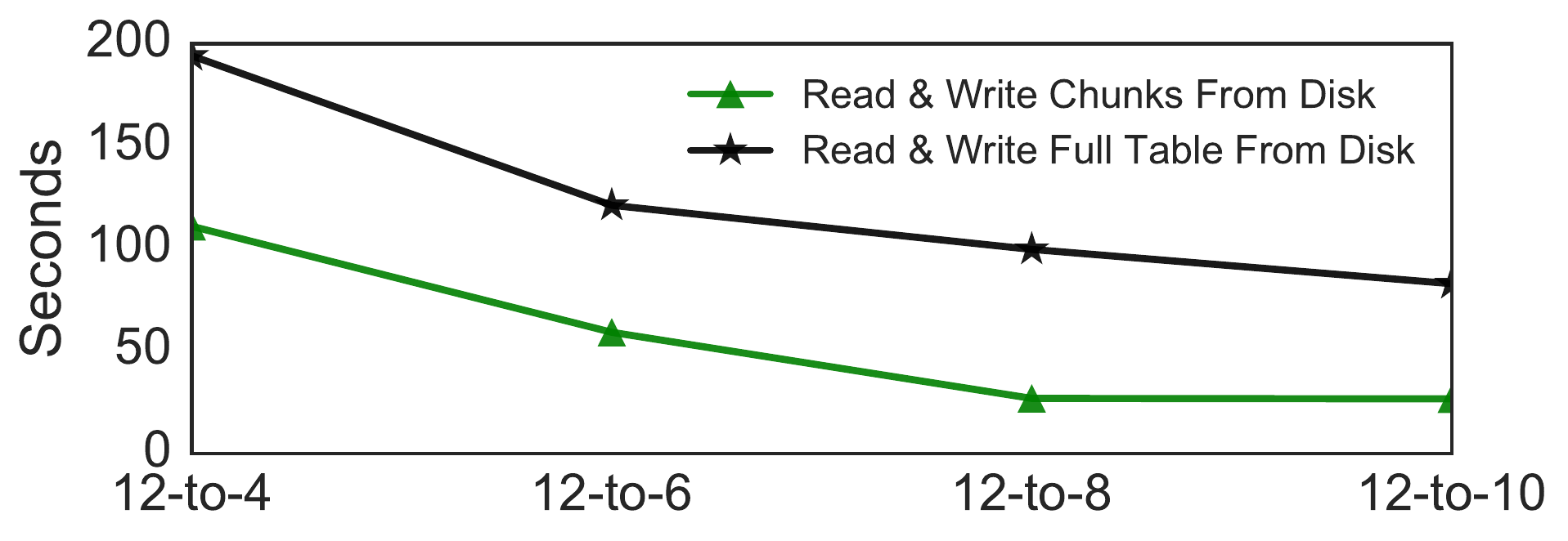}
  
  \caption{Time to Change Configurations from 12 Workers}
  \label{fig:reconfiguration_12Workers}
  \end{subfigure}
  \caption{Configuration Change Runtimes}
  \vspace{-0.7cm}
\end{figure}

Shuffled-Scaling must re-organize the data before running the next query. In \autoref{fig:reconfiguration_4Workers}, we show the amount of time it takes for a configuration of 4 workers to move data to a configuration $c'$. \autoref{fig:reconfiguration_12Workers} shows the amount of time it takes to change from a set of 12 worker nodes to $c'$. We evaluate two approaches to switch between configurations for Shuffled-Scaling. As a first method, \name reads the entire table from disk in $c$, shuffles the data, and writes it to $c'$. We call this approach \textit{Read \& Write Full Table from Disk}. As a second method, the system uses the optimization described above that only reads and writes the minimum amount of data, which we denote with \textit{Read \& Write Chunks From Disk}. The longest configuration switch is from 12-to-4, which implies that the bottleneck is in the time it takes to write data to disk. In all cases, however, data reshuffling creates a visible interruption in the query session.

\begin{figure}[t]
  \centering
  \includegraphics[width=.8\linewidth]{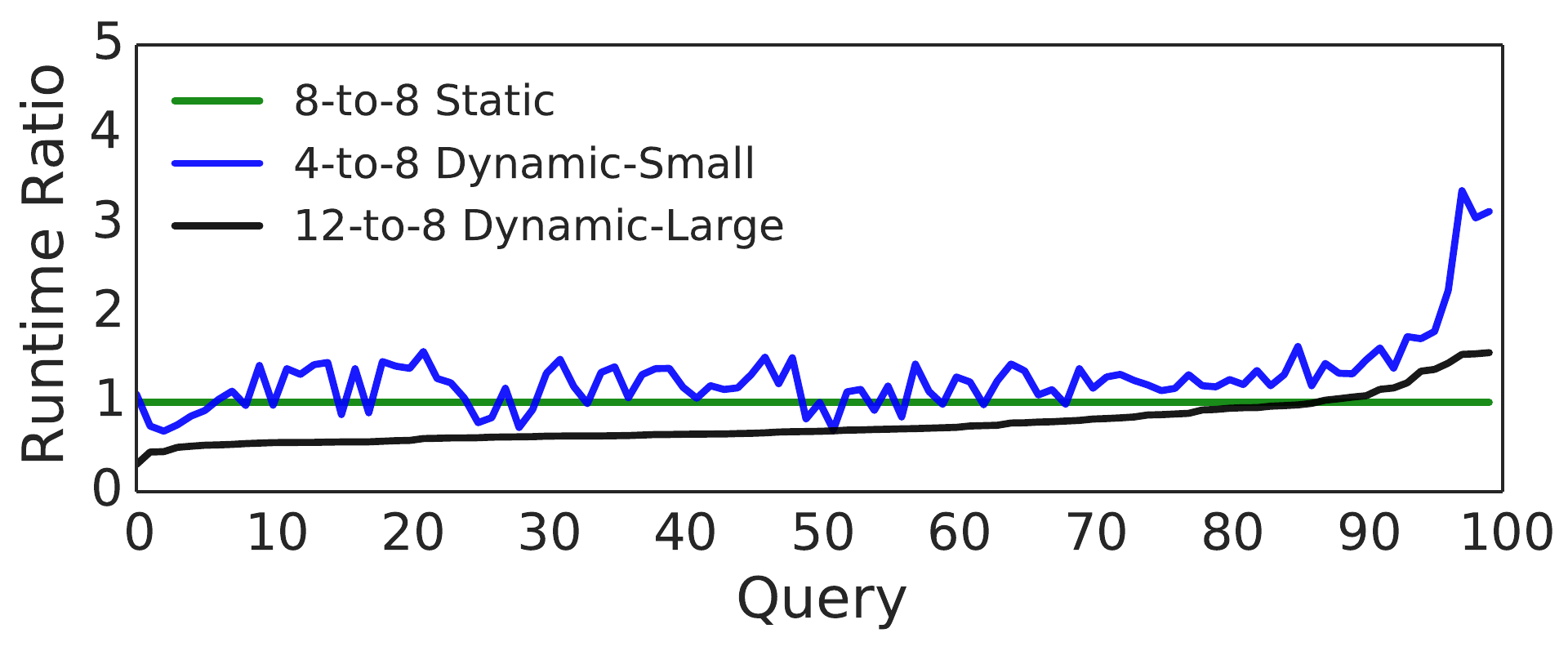}
  \caption{Time to run a random set of TPC-H SSB queries in static and dynamic clusters}
  \label{fig:static_dynamic_runtimes}
  \vspace{-0.6cm}
\end{figure}

\textbf{Query Processing Time} So far, we showed that Shuffled-Scaling imposes too much overhead during cluster resizing while Static-Replicated can take a long time to ingest data. Here, we compare the remaining competitive methods on their query execution times.  \autoref{fig:static_dynamic_runtimes} shows the query runtime ratios for 100 randomly selected queries from our generated pool of queries.  The ratios measure the query time for Dynamic-Small (4 data and 8 compute nodes) and Dynamic-Large (12 data and 8 compute nodes) compared with Static-Replicated Chunks (8 nodes shown as 8-to-8 static in the figure). In this experiment, we only measure the computation time for each query and do not flush the query results to disk. As the figure shows, Dynamic-Small leads to slower query runtimes than the Static-Replicated method. In general, we observe in our experiments (results not shown due to space constraints) that a small set of data nodes can easily become a bottleneck and nullify any benefit of scaling compute nodes. In contrast, Dynamic-Large has excellent performance. This latter method, however, is expensive. We find that it delivers high performance only when data nodes use powerful VMs. When using cheaper nodes, the data nodes become a bottleneck again (results not shown). 
However, if a data node uses a powerful VM, it has the capability to also run as as a compute node. Shuffled-Scaling, Static-Replicated and Static-Replicated Chunks already co-locate compute and data nodes, and thus are more cost effective compared to dynamic methods. 






\textbf {Cost of Virtual Machines} At the start of the query session, \name launches the number of VMs necessary to meet all configuration options in $configs$. \name can help minimize $CS(Q)$ without significantly penalizing query performance by turning off VMs that are not in use. The time it takes to launch a new virtual machine in \name depends on the size of the virtual machine. For an m3.large machine, it takes approximately 17 seconds to launch a machine with an Amazon Linux AMI. Turning off the machine takes 27 seconds on average. Turning a machine back on takes only approximately 10 seconds.  


\textbf{Data Organization Summary} Although Dynamic-Scaling can have the lowest data ingest times, this option is costly and risks slowing down query processing if an insufficient number of data nodes are selected as shown in \autoref{fig:static_dynamic_runtimes}. Given this result, the best option is for \name to use the Static-Replicated Chunks scaling method as it provides a quick way to ingest data and does not incur any runtime penalties when switching between configurations. 



\vspace{-0.25cm}
\section{Scaling Algorithms} \label{sec:scaling_algorithms}

In this section, we consider both reactive and proactive methods for
PerfEnforce to rescale the user's cluster during her query session.
We introduced these methods initially in a short, demonstration
proposal~\cite{Ortiz:16}. The contribution of this paper lies in the
actual study of these methods. The goal of these scaling methods is to
maintain $PR(Q)$ as close to 1.0 as possible.

\subsection{Reactive Scaling Algorithms}
\label{sec:scaling:reactive}

We first describe reactive scaling algorithms. These algorithms take
action after they witness either a good or bad event. In \name, we
implement proportional integral control and reinforcement learning as
our reactive methods because these methods have successfully been
used in other resource allocation contexts~\cite{Lim:10,Konstantinou:12}.

\textbf{Proportional Integral Control (PI)}  Feedback control~\cite{Janert:13} is a commonly used approach to regulate a system in order to ensure that it operates at a given reference point.  We use a proportional-integral controller (PI) as a method that helps \name react based on the magnitude of the error while avoiding oscillations over time.


At each time step, $t$, the controller produces an actuator value
$u(t)$ that causes the system to produce an output $y(t+1)$ at the
next time step. The goal is for the system output $y(t)$ to be equal
to some desired reference output $r(t)$. In an integral controller,
the actuator value depends on the accumulation of past errors of the
system. This can be represented as $u(t+1) = u(t) + k_i e(t)$. Where $e(t) = y(t) - r(t)$, with $y(t)$ being the observed output and $r(t)$ being the target system output. $k_i$ represents the gain of the integral control. Ideally, this parameter is tuned in such a way that helps drive $e(t)$ to 0. In our scenario, the actuator value $u(t)$ is the discrete number of VMs provisioned. 

As for the system output, $y(t)$, we use the average ratio of the real
query runtime $t_{real}(q)$ over the query runtime promised in the
SLA, $t_{sla}(q)$, over some time window of queries $w$ as $y(t) = \frac{1}{|w|} \sum_{q \in w} \frac{t_{real}(q)}{t_{sla}(q)}$ where $|w|$ is the number of queries in $w$. 


Our target operating point is thus $r(t) = 1.0$ and the error $e(t) = y(t)-r(t)$ captures a percent error between the current and desired average runtime ratios. Since the number of VMs to spin up and remove given such a percent error depends on the cluster size, we add that size to the error computation as follows: $e(t) = (y(t) - r(t))u(t)$.

Integral control alone may be slow to react to changes in the workload.
Therefore, we also introduce a proportional control component, where
$k_p$ represents the gain of the proportional error.
Our final PI controller thus takes the following form:

\begin{equation} \label{eq:pi-control}
u(t+1) = u(0) + \sum_{x=0}^{t} k_i e(x) + k_p e(t)
\end{equation}

\textbf{Reinforcement Learning (RL)} As our second reactive method, we use reinforcement learning (RL). This approach has successfully been applied in the TIRAMOLA system, which supports elastic scaling of NoSQL databases~\cite{Konstantinou:12}.


At each state $s$, the model makes a probabilistic decision to move to
another state $s'$ by taking an action $a$. In our case, each state
represents a configuration in $configs$ and the action is to change to
that configuration. The goal is to make a series of beneficial
action-state moves, as motivated by the rewards at each state,
$R(s)$. To explore the search space and
learn the optimal action-state paths, reinforcement learning uses a
technique known as Q-learning \cite{Sutton:98}:

\begin{equation} \label{eq:Q_learning}
 Q(s,a) = Q(s,a) + \alpha[R(s') + \gamma \underset{a}{\text{max}} Q(s',a')- Q(s,a)]
 \end{equation}

 In \autoref{eq:Q_learning}, Q(s,a) is the reward
 for taking action $a$ from state $s$. It is a function of the reward
 at state $s'$ reached by taking action $a$ and of the actions
 that can subsequently be taken from $s'$. $\alpha$ represents
the learning rate, which controls how fast the learning takes
place. At convergence, Q-learning is able to find an optimal
action-state path. In \name, our goal is different. Since a user's
query workload is constantly changing throughout the session,
recording the action-state path is unnecessary. Instead, \name directly
transitions to the state with the highest reward. We define the reward
function to be the real-to-SLA runtime ratio. At each iteration, we
favor states with the reward closest to 1.0, where the real query
runtimes are closest to the SLA runtime. As the system transitions to
a state $s$, it updates the reward function for that state. We use the
following equation, where $R(s)$ denotes the updated reward for state
$s$:

\begin{equation} \label{eq:RL_update}
  R(s) = \alpha * (\frac{t_{real}(q)}{t_{sla}(q)} - R(s)) + R(s)
\end{equation}

At the initialization of the model, each state must begin with a
defined reward value, $R(s)$. This implies that the system must have
prior knowledge of the performance of the user's queries for each
configuration. Since we do not have such prior knowledge, we set the
reward at each state to 1.0 and force the system to first explore
states that are closest to $init_c$. To do this, we maintain a set of
states called \textit{active states}. When the query session begins,
\textit{active states} only contains the configuration $init_c$. If
the reward for the current state goes above 1.0, we add the next
larger cluster size to the active states. If the reward for the
current state goes below 1.0, we similarly add the next smaller
cluster size. We repeat the process until all possible cluster sizes
have been added.


Additionally, we observe that rewards for some states do not
quickly adapt if the user's workload changes. For example, if a slow query
runs on configuration $c$ and misses the deadline, the reward will be
updated to a value above 1.0. If a new fast query is introduced, $c$
will not be chosen as the current reward (above 1.0) suggests that the
query will miss the deadline. Therefore, as a heuristic, we introduce
a linear-drag update. Each state whose reward was not modified by
\autoref{eq:RL_update} (denoted as state $x$), receives the following update:
$R(x) = \beta * (\frac{t_{real}(q)}{t_{sla}(q)}*\frac{y}{z} - R(x)) +
R(x)$. Where $\beta < \alpha$ and $z$ represents number of VMs of state $x$.
$y$ is the number of VMs in state $s$ from \autoref{eq:RL_update}.

\subsection{Proactive Scaling Algorithms}
\label{sec:scaling:proactive}

Instead of approaches that react to runtime errors such as PI and RL,
we also explore an approach that makes use of a predictive model. For
each incoming query, \name predicts the runtime for the query for each
configuration and switches to the closest configuration where the ratio
is closest to 1.0.

\name first builds an offline model for a given cloud data analytics
service. For training data, we use the Parallel Data Generation
Framework tool~\cite{Rabl:10} to generate a 10GB dataset with a set of
6120 queries. These training queries are based on the query generator
provided by the open-source PSLAManager tool~\cite{ortiz:15}.
Training data consists of query plan features including the estimated
max cost, estimated number of rows, estimated width, and number of
workers.


Initially, such an offline model is expected to be inaccurate.
However, we can adaptively improve the model if we incorporate
information about the queries the user executes on his data. We
achieve this goal by using a perceptron learning model: as the user
executes queries, \name improves the model in an online fashion. We
use the MOA (Massive Online Analysis) tool for learning~\cite{moa}.

\textbf{Perceptron Online Machine Learning (OML)}  The perceptron learning algorithm works by adjusting weights for each new data point. We find that it adapts more quickly to new information than an active-learning based approach. \name initiates the perceptron model by first learning from the training set. For an incoming query, \name uses this model to predict a runtime for each configuration in $configs$. The cluster size with the closest runtime to the incoming query's SLA is chosen. Once the system runs the query and learns about the real runtime, it feeds this information back into the model. If we predict in parallel for all $configs$, the process takes less than 1 second.

\vspace{-0.25cm}
\section{Scaling Algorithms \\ Evaluation} \label{sec:scaling_algorithms_eval}

We now evaluate both reactive and proactive scaling algorithms. We first execute a series of microbenchmarks designed to demonstrate fundamental characteristics of each algorithm. We then evaluate each algorithm on macrobenchmarks consisting of random workloads. We keep the same experimental setting as before where we have $configs = \{4,6,8,10,12\}$ and the TPC-H SSB dataset. We run queries from our query pool for each cluster configuration and record the execution times. We label each query with the cluster configuration that is able to run the query at the runtime closest to the query's SLA.  For all experiments, we set $init_c$ to 4.

\subsection{Scaling Algorithms Microbenchmarks} \label{sec:micro}

\begin{figure*}
  \centering
  \begin{subfigure}{.9\textwidth}
  \centering
  \includegraphics[width=\linewidth]{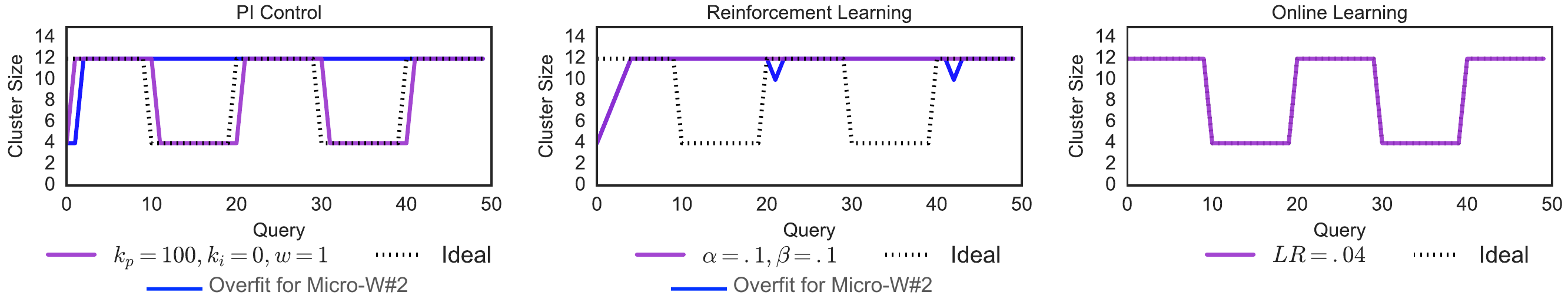}
  \caption{Micro-Workload \#1: Convergence Speed}
  \label{fig:convergence}
  \end{subfigure}
  ~
  \centering
  \begin{subfigure}{.9\textwidth}
  \centering
  \includegraphics[width=\linewidth]{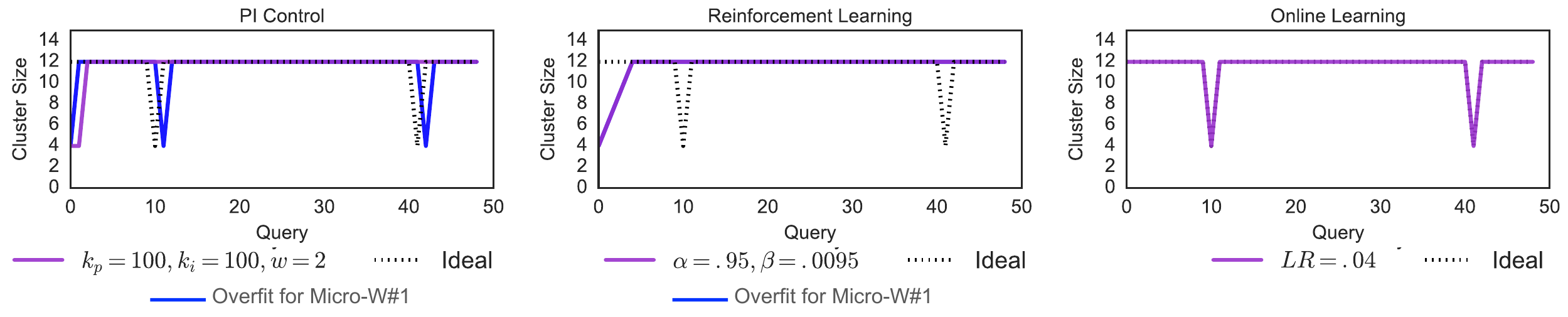}
  \caption{Micro-Workload \#2: Stability in the Face of one Different Query}
  \label{fig:stability}
  \end{subfigure}
    ~
  \centering
  \begin{subfigure}{.9\textwidth}
  \centering
  \includegraphics[width=\linewidth]{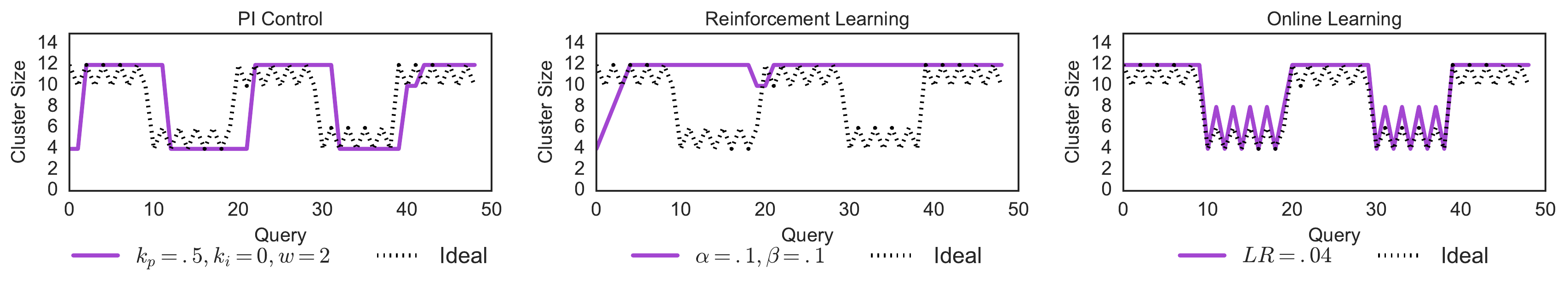}
  \caption{Micro-Workload \#3: Tracking of a Rapidly Changing Workload}
  \label{fig:besteffort}
  \end{subfigure}
  \caption{Micro-Benchmarks}
  \vspace{-0.5cm}
\end{figure*}

For each algorithm, we consider the following: (1) How fast does the algorithm converge to a different configuration if the current configuration is either too small to meet the SLA times or is unnecessarily too large? How fast does the algorithm react to a workload change that requires a different cluster size? (2) How stable is the algorithm in the face of occasional queries that would require either a smaller or larger cluster than the rest? (3) How well does the algorithm handle an oscillating workload where the ideal cluster size is different for consecutive queries? We use the following three workloads to help answer these three questions: (1) \textbf{Micro-W\#1}: Convergence Speed. (2) \textbf{Micro-W\#2}: Stability.  (3) \textbf{Micro-W\#3}: Workload tracking.  

All three scaling algorithms have tuning parameters. In this section, we show the performance of reinforcement learning (RL) and PI-control (PI) with best parameters chosen \textit{separately} for each workload. The selected parameters \textit{overfit} the workload. We describe how we select the best parameters for PI and RL in \autoref{sec:macro}. For perceptron online machine learning (OML), we select a learning rate of $0.04$. In \autoref{sec:oml_tuning}, we discuss how to tune OML. For Micro-W\#1 and Micro-W\#2, we also include an additional $overfit$ line for a different workload (shown in blue). The goal is to show how tuning parameters for one workload do not necessarily benefit scaling for other workloads. We do not show this for OML since we use the same parameter value for all workloads. 


\textbf{Micro Workload 1- Convergence Speed} In this first workload, we evaluate the speed of convergence for each technique on the workload shown in \autoref{fig:convergence}. The system starts at $init_c$, a 4-worker configuration. The query sequence begins with a set of 10 queries whose SLA deadline is best met at 12 workers. This is then followed by a set queries whose SLA is more closely met at 4 workers. We repeat this pattern for a total of 50 queries.  For PI, the model immediately scales to the largest cluster size after running the first query in the sequence. PI is able to converge because we tune the PI controller to react quickly with a $k_p$ value of 100. Although $k_p$ is high, it does not oscillate once it converges to 12 workers since $e(t)$ turns out to be positive for each of these queries (recall, we initially select queries whose SLA is best met at 12 workers, but might not necessarily meet the guarantee at this configuration size).  Nevertheless, there exists a lag between the workload change and the PI's reaction to that change. In contrast, OML is able to track the workload exactly as it correctly predicts the required cluster size.  For RL, scaling does not happen as quickly as seen in PI, as states are incrementally added to the \textit{activeStates} set. Therefore, convergence for the first set of queries does not occur until the 5th query. Linear-drag updates still take place in this workload, but the sequences between slow and fast queries are not long enough to be able to see its effect. As a result, RL remains in one state.

\textbf{Micro Workload 2- Stability} For this workload, we show the stability of the scaling algorithms in situations where the system runs a fast query among a long sequence of queries whose guarantee is best met at 12 workers. The results are shown in \autoref{fig:stability}. For the PI controller, the best settings are those where $k_i$ or $w$ value is large. With these settings, the PI controller is stable in face of outlier queries. Observe, however, that the PI controller settings are now different compared to W\#1. As we still show in  \autoref{sec:tuning_search_space}, the PI controller is highly sensitive to its parameter settings. For RL, we see a similar behavior. The ideal setting uses a high $\alpha$ parameter, the rewards for configurations running the first few queries are updated to a high ratio, since they all miss the SLA deadline. This then makes it difficult for the model to quickly scale back down later in the sequence. Once the fast query runs at 12 workers, the model updates the reward for this configuration, but this state continues to be the closest to 1.0. OML is able to determine the ideal cluster configurations before running each query. In general, a one-time-only query with a different ideal cluster size does not negatively impact the result for any method. In this example, PI and RL only end up over provisioning for these fast queries. 

\textbf{Micro Workload 3 - Workload tracking} Finally, we demonstrate how well each method is able to keep up with a rapidly changing workload. We demonstrate this through a sawtooth workload where we first run a mix of queries whose SLA is best met at 12 and 10 workers, followed by a mix of queries whose SLA is met at 4 and 6 workers as shown in \autoref{fig:besteffort}. For PI, one of the parameter combinations that work best for this workload are $k_p=.5, k_i=0, w=2$. The model immediately scales up after the first window of queries. Similarly to the first workload, PI continues to scale by simply reacting to the error due to the $k_p$ parameter.  RL scales to the highest cluster size as before. It temporarily scales down to a configuration of 10 queries thanks to linear-drag, but the model quickly scales back up since it under-provisions the 20th query. In OML, the model over provisions for several queries, by at most one configuration size.

In general, reactive methods are able to converge and even recover in the presence of a sudden change in the workloads. However, they are difficult to tune especially for rapidly changing workloads. OML is able to keep up with rapidly changing workloads given that it is able to observe the features for the upcoming query before choosing the best cluster configuration. 

\vspace{-0.25cm}
\subsection{Perceptron Learning Tuning} \label{sec:oml_tuning}

We now discuss how to find an optimal learning rate for OML. Recall, the learning rate for OML determines how quickly the model adjusts the weights for different features. If the learning rate value is too low, the model might not quickly adapt to the user's new queries. If it is high, learning is faster, but there is a risk that the model might never converge, as it will tend to jump over the optimum. To evaluate the sensitivity, we first take three random sets of 100 queries from the TPC-H SSB dataset. For each of these test sets, we also prepare a separate group of 400 holdout queries from the same dataset. For offline training, we use a total of 6120 queries from the Parallel Data Generation Framework dataset \cite{Rabl:10}. 

\begin{figure}
\centering
\includegraphics[width=.9\linewidth]{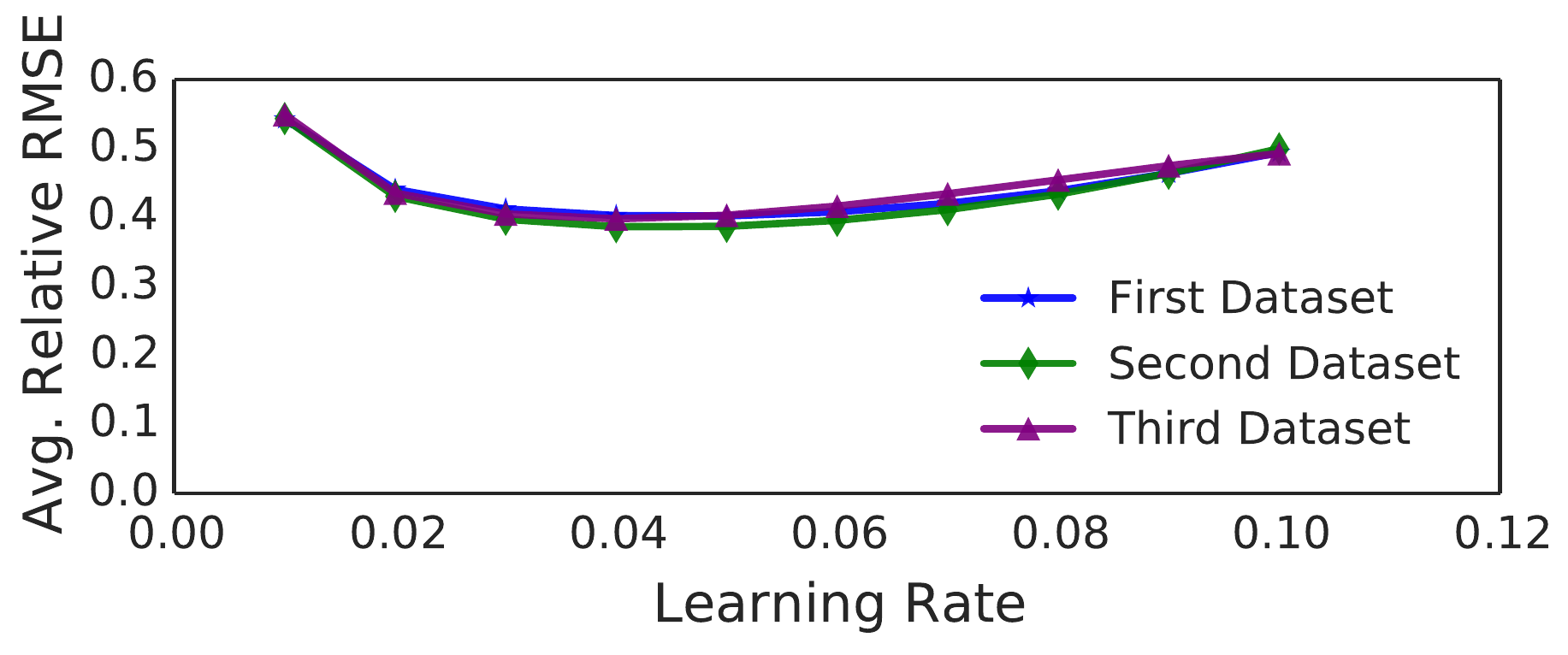}
\caption{Tuning for OML based on the TPC-H SSB Dataset}
\label{fig:oml_tuning}
\vspace{-0.5cm}
\end{figure}

We first select a learning rate and a test set. For each query in the test set, we add it to the training model, update the model, and evaluate the model on the corresponding holdout set of queries to collect the relative root mean squared error (relative RMSE). Once we evaluate against all the queries in the test set, we calculate the average relative RMSE. We repeat this process for many learning rates. \autoref{fig:oml_tuning} shows the resulting average relative RMSE on the y-axis for different learning rates on the x-axis. As the figure shows, for all datasets, the learning rate with the lowest average relative RMSE is approximately $0.04$, which is the value that we use in all other experiments in this section. Importantly, however, a large range of learning rates $[0.02,0.07]$ yield similar prediction quality. 


\subsubsection{Effects of Caching and Contention}

In the previous section, tuned for OML based on a cold-cache environment. We evaluated a cold-cache set of test queries against a cold-cache training model, $c_{train}$. In a real query session, \name will not clear the cache after each query runs. 

\begin{figure}[t]
  \begin{subfigure}{.5\textwidth}
  \centering
  \includegraphics[width=.8\linewidth]{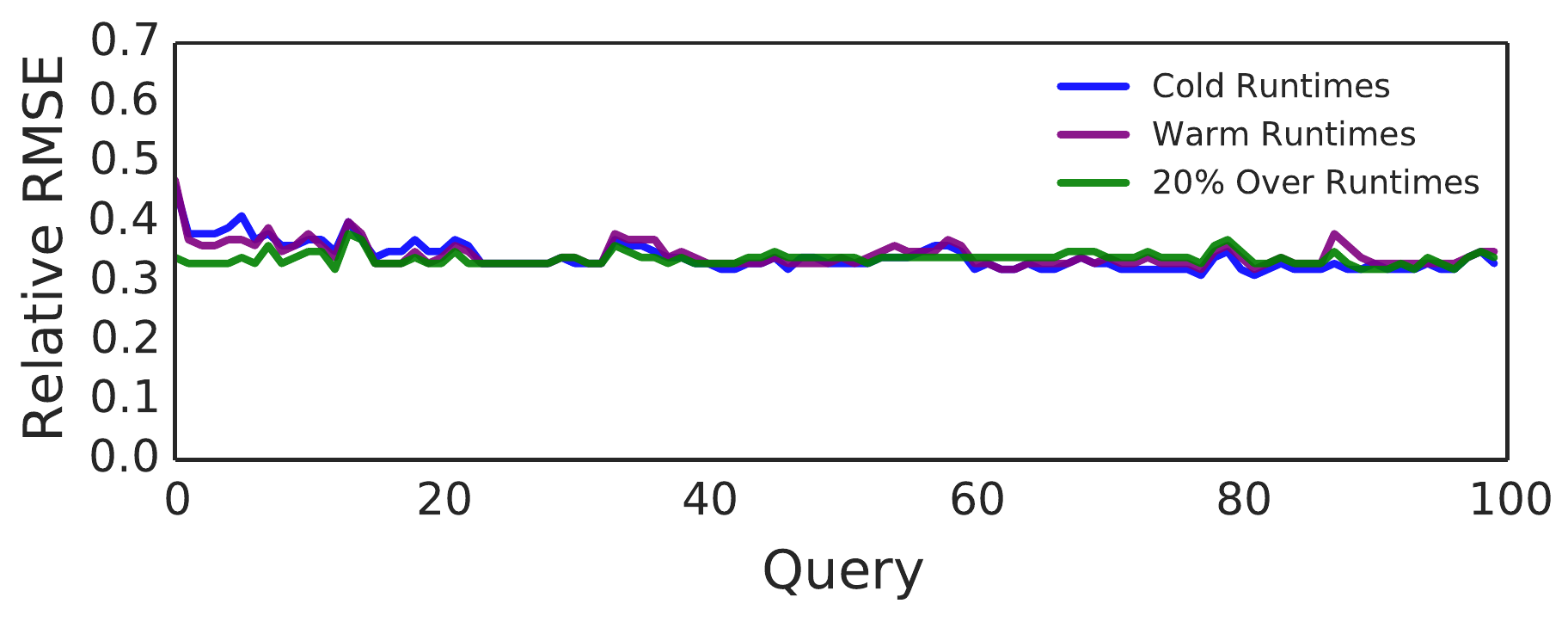}
  \caption{Prediction Errors based on $c_{train}$}
  \label{fig:cold_training}
  \end{subfigure}
  ~
  \begin{subfigure}{.5\textwidth}
  \centering
  \includegraphics[width=.8\linewidth]{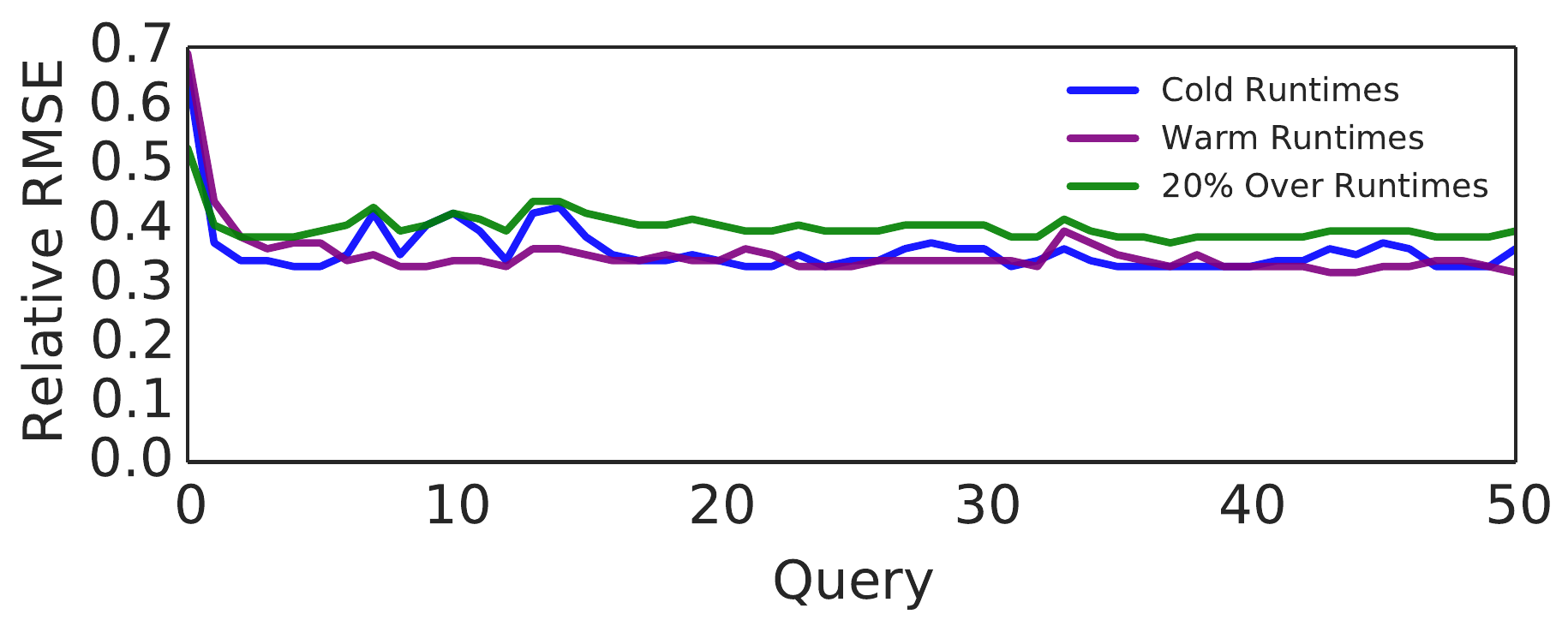}
  \caption{Prediction Errors based on $w_{train}$}
  \label{fig:warm_training}
  \end{subfigure}
  ~
  \caption{Prediction Errors between $c_{train}$ and $w_{train}$}
  \vspace{-0.7cm}
\end{figure}

In practice, queries should execute faster in a warm environment. We briefly evaluate the effects of data caching with query runtime predictions. We first generate a warm-cache training model, $w_{train}$ in order to observe if the $w_{train}$ model outperforms the $c_{train}$ model when it comes to predicting the runtimes in a warm query session. To record the runtimes for the offline warm training and testing models, we run each query twice and only record the runtime of the second query. \autoref{fig:cold_training} shows the prediction error for a set of 100 queries. We evaluate $c_{train}$ against three versions of the test queries: cold-cache runtimes, warm-cache runtimes and with a 20\% additional time (to model contention) above the cold-cache runtime. In general, $c_{train}$ is able to improve the predictions over time for all three sets of queries. However, $w_{train}$ does not perform as well. As shown in \autoref{fig:warm_training}, the error starts off higher than $c_{train}$ at approximately a Relative RMSE of $0.7$. Second, although $w_{train}$ is able to achieve a low Relative RMSE for warm test runtimes, there is more significant error for cold test runtimes.  Overall, an offline model trained on cold-cache queries is more resilient and can adapt to predicting either cold-cache or warm-cache query times.





\vspace{-0.25cm}
\subsection{Scaling Algorithms \\ Macrobenchmarks} \label{sec:macro}

In this section, we focus on the performance of the scaling algorithms on random workloads. We seek to answer two questions: (1) On random workloads, how well do different techniques manage to operate at the desired $PR(Q)=1$? (2) What is the cost of operating at the given set point?


For RL and PI, we first show the performance when selecting the best parameter settings separately for each workload. We call this variant \textit{overfitted}, since the parameters are completely overfitted to the workload and thus change for each sequence of queries. To find overfit parameters, we use information from an \textit{Oracle}. The Oracle is an additional technique that holds all knowledge of queries and their corresponding ideal configurations. The Oracle executes the same workload of queries, picking the best cluster configuration for each query. We compute the $PR(Q)$ for the workload as executed by the Oracle. We then iterate through all possible combinations of parameters for each technique. For RL, we iterate through $\alpha$ rates from 0 to 1.0 where $\beta = \frac{\alpha}{d}$ and we vary $d$ from 1 to 100.  For PI, we vary $k_p$ and $k_i$ values from 0 to 100, with varying window sizes, $w$, from 1 to 100. For each parameter combination, we execute the technique on the given workload. We show the results for the parameter combination that yields a $PR(Q)$, closest to that of the Oracle. For OML, we continue to use the optimal learning rate of $0.04$.


\begin{figure*}[t]
 \centering
 \begin{subfigure}{.9\textwidth}
 \includegraphics[width=\linewidth]{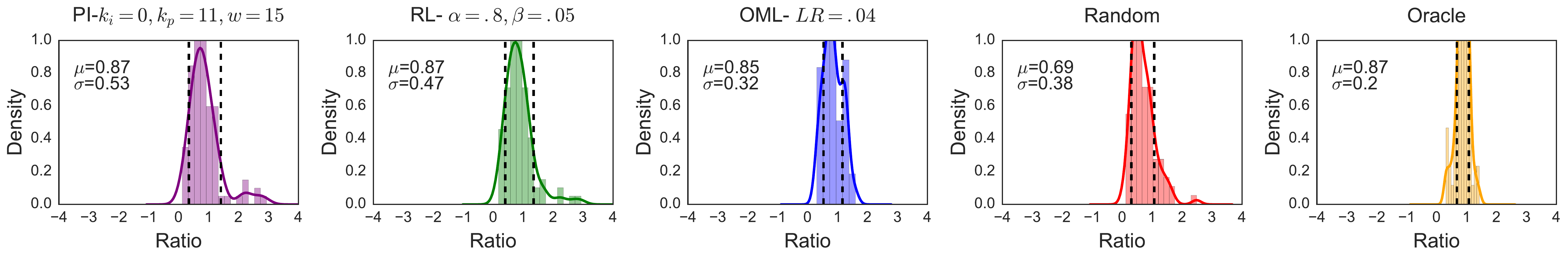}
 \caption{Workload: With Convergence at 4 Workers}
 \label{fig:random_true}
 \end{subfigure}
 ~
 \begin{subfigure}{.9\textwidth}
 \centering
 \includegraphics[width=\linewidth]{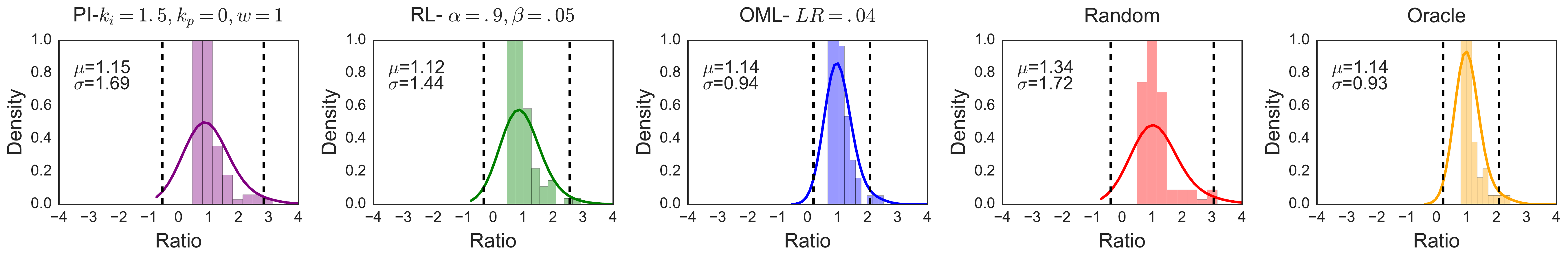}
 \caption{Workload: No-Convergence}
  \label{fig:random_bad}
 \end{subfigure}
 ~
 \caption{\textbf{Ratio Distributions of Random Workloads}: Distributions of $\frac{t_{real}(q)}{t_{sla}(q)}$ ratios for each technique on two random workloads. RL and PI use overfitted parameters. OML uses a learning rate of .04}
\label{fig:dist}
 \vspace{-0.5cm}
\end{figure*}

\subsubsection{Ratio Distributions for Random Workloads }

We first show the distributions of query runtime ratios for two concrete random workloads. Each workload comprises 100 random queries from our set of TPC-H SSB queries. 

\autoref{fig:random_true} shows the ratio distribution of each technique for the first random workload. The x-axis shows the, $\frac{t_{real}(q)}{t_{sla}(q)}$ ratio. The y-axis shows the density for each ratio (\ie, the fraction of queries with that ratio). In addition to the Oracle's distribution, we also show the distribution for a \textit{random} technique. The random technique simply selects a random cluster size to run each query. As the figure shows, the Oracle's distribution has an average of $0.87$, which implies that there are many queries in the workload where the smallest cluster size available in $configs$ does not closely meet the query's SLA. The system could scale down further, but it does not because we set the lower limit at four workers.


For PI and RL, we show the distributions based on the overfit parameters. Both RL and PI run queries that at times are 2x or 3x slower than the query's SLA guarantee. OML is able to follow the Oracle's distribution more closely with most of the queries falling between the ratios $0.53$ and $1.17$. We find that for this random workload in particular, many of queries are able to meet their SLA at a 4 worker configuration. This provides an opportunity for RL and PI to converge to this configuration size for a majority of the queries. We refer to this workload as the \textit{Convergence} workload. For this workload, all techniques achieve a $PR(Q)$ close to that of the Oracle. However, the standard deviation is much larger for PI and RL than for OML.

For the second random workload, we only select queries whose SLA is not met at 4 workers. We refer to this workload as \textit{No-Convergence}. We show the distributions for No-Convergence in \autoref{fig:random_bad}. The Oracle distribution has a higher average and standard deviation for this workload. This is due to some queries in the workload not being able to meet their SLAs even at the largest configuration size. The overfit parameters for PI and RL are not as close to the Oracle distribution as for the previous workload. Standard deviations are even higher. For both of these techniques, there are queries that run up to 3x slower than their assigned SLA runtime. OML is able to produce a similar distribution as the Oracle.

\begin{figure}[t]
 \begin{subfigure}{.5\textwidth}
 \centering
 \includegraphics[width=.9\linewidth]{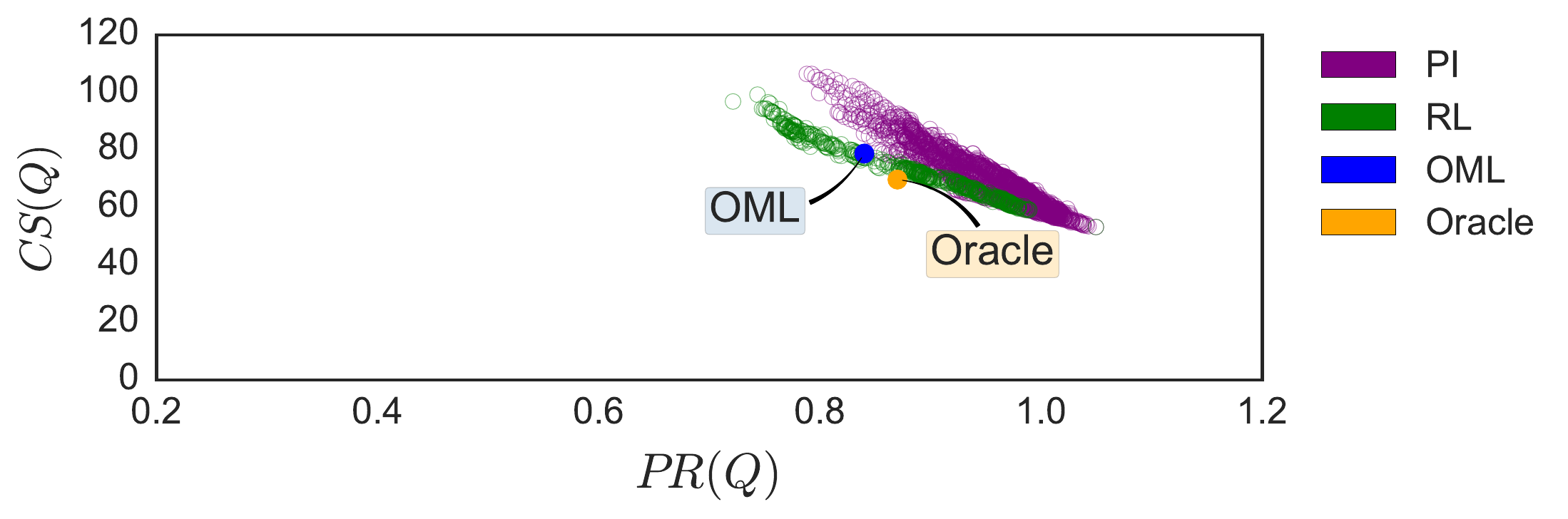}
 \caption{Parameter Sensitivity for Convergence Workload}
 \label{fig:search_space_true}
 \end{subfigure}
 ~
 \begin{subfigure}{.5\textwidth}
 \centering
 \includegraphics[width=.9\linewidth]{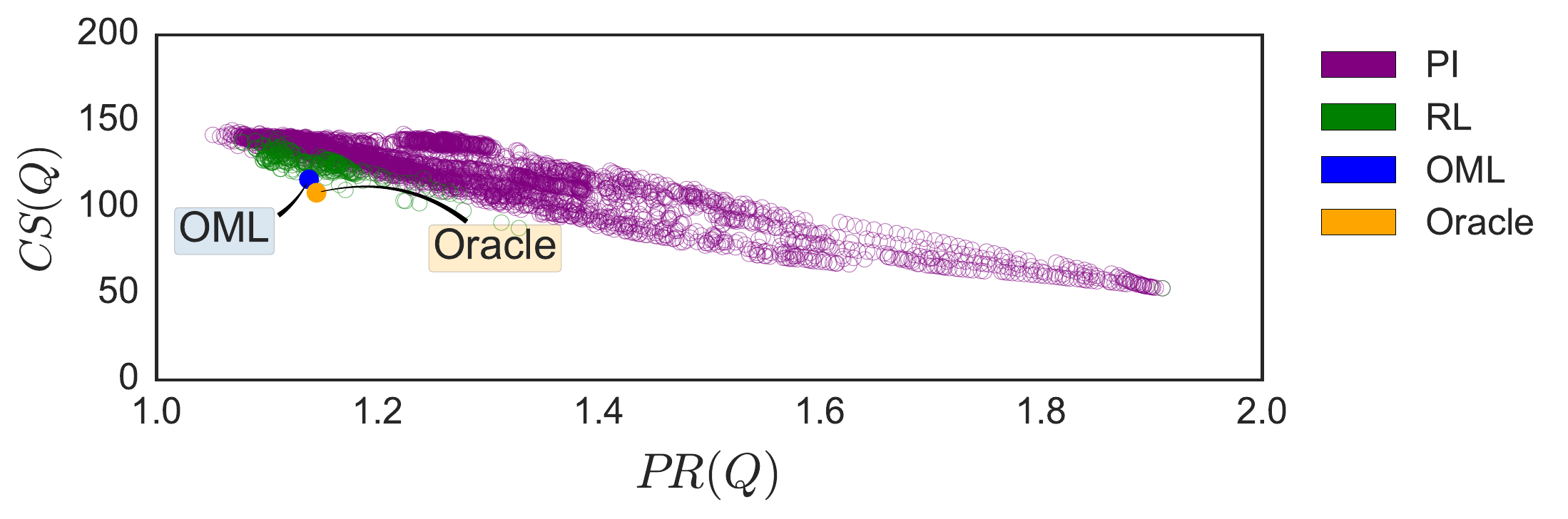}
 \caption{Parameter Sensitivity for No-Convergence Workload}
 \label{fig:search_space_false}
 \end{subfigure}
 ~
 \caption{Parameter Value Sensitivity: Each point represents an execution with different parameter values.}
 \vspace{-0.6cm}
\end{figure}

\begin{figure*}
\centering
\includegraphics[width=\linewidth]{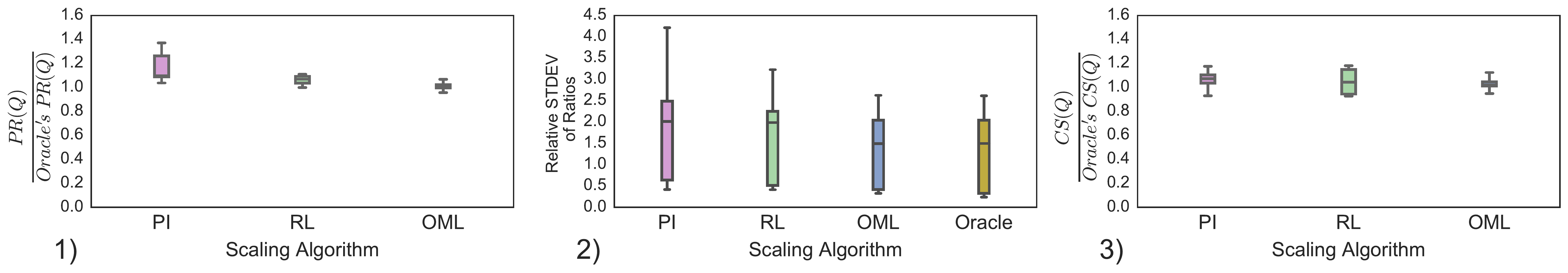}
\caption{Performance for Techniques across Many Workloads using Average Parameter Settings}
\label{fig:average_parameters}
\vspace{-0.7cm}
\end{figure*}

\subsubsection{The Parameter Search Space} \label{sec:tuning_search_space}

For the Convergence and No-Convergence workloads, we previously showed only the distributions for overfit parameter values. We now show a summary of the distributions for different parameter values. \autoref{fig:search_space_true} and \autoref{fig:search_space_false} show $PR(Q)$ \textit{vs.} $CS(Q)$ for the resulting distributions. We also included the resulting distribution for OML (based on a learning rate of .04) and the Oracle.  As the figure shows, both PI and RL are highly sensitive to their parameter settings. Wrong settings can yield high query time ratios or high costs (measured in terms of virtual machines).  Additionally, we find that the best settings vary significantly across workloads. These techniques are thus impractical in our setting, where the workload is unknown until the user starts issuing queries. 



\subsubsection{Performance for Many Random Workloads}


We evaluate each of the techniques on a set of ten workloads. Five of those workloads are completely random, two workloads are random but comprise only large queries without selection predicates, one workload comprises only queries that select 10\% of the data, and the final two workloads either comprise a majority of queries that have an ideal cluster of size 4 or an ideal cluster of size 12.  For OML, we continue to use the same learning rate for each workload. For PI and RL, we first find the overfit parameter values for each workload and then compute an overall average value for each parameter. We show the performance of the techniques on the average settings, since in practice, the system cannot predict the workload and set the optimal parameter values for each workload.




\autoref{fig:average_parameters} shows the results of using these average parameter values. In \autoref{fig:average_parameters} (1), we show the distribution of $PR(Q)$ across the ten workloads for each technique relative to the Oracle's $PR(Q)$. In \autoref{fig:average_parameters} (2) we show the distribution of the relative standard deviations across the workloads for each technique. The relative standard deviation is taken by calculating the standard deviation of the resulting query ratios ($\frac{t_{real}(q_i)}{t_{sla}(q_i)}$) for each workload and dividing it by the mean. Finally, \autoref{fig:average_parameters} (3) shows the distribution of the cost of service ($CS(Q)$) across all workloads for each technique relative to the Oracle's $CS(Q)$. As the figures show, OML yields ratio distributions closest to those of the Oracle: Both the $PR(Q)$ and relative standard deviations are closer to those of the Oracle compared with PI and RL. Both PI and OML yield similar $CS(Q)$ to the Oracle.

\vspace{-0.25cm}

\subsection{From QoS to SLA}

In the previous sections, we evaluated the elastic scaling algorithms in terms of how well they enable the system to operate close to the desired set point where $PR(Q)=1$.  In this section, we discuss how the results can translate into a concrete SLA. As we showed above, each cluster scaling algorithm produces a distribution of these ratios around the desired set point and we showed that OML yields a distribution close to that of the Oracle without difficult parameter tuning. In \autoref{tab:percentiles}, we show the resulting query time ratios for different percentiles in the distributions obtained for the 10 random workloads. As the table shows, the ratios are consistent across the workloads and are relatively close to the desired set point. A cloud provider can thus use PerfEnforce with the OML scaling algorithm and advertise a probabilistic SLA, where the cloud provider increases the estimated query runtimes by some weight $w$ and then promises that $X\%$ of the queries will meet their SLA runtimes. That is, the SLA will promise fewer than $1-X\%$ SLA violations in a session. Considering the table, possible SLAs include advertising query times that are $w=2$ times slower than actually anticipated and offering fewer than $10\%$ SLA violations. Another option would be $w=1.7$ with fewer than $15\%$ SLA violations in a session. Depending on SLA violation costs, of course, the cloud may choose to be more or less conservative.





\begin{table}[t]
\centering
\scalebox{0.7}{
\begin{tabular}{l|llllllllll}
\cline{2-11}
                                     & \multicolumn{1}{l|}{\textbf{W\#1}} & \multicolumn{1}{l|}{\textbf{W\#2}} & \multicolumn{1}{l|}{\textbf{W\#3}} & \multicolumn{1}{l|}{\textbf{W\#4}} & \multicolumn{1}{l|}{\textbf{W\#5}} & \multicolumn{1}{l|}{\textbf{W\#6}} & \multicolumn{1}{l|}{\textbf{W\#7}} & \multicolumn{1}{l|}{\textbf{W\#8}} & \multicolumn{1}{l|}{\textbf{W\#9}} & \multicolumn{1}{l|}{\textbf{W\#10}} \\ \hline
 \multicolumn{1}{|l|}{\textbf{75\%}} & .82                                & 1.21                               & .83                                & .93                                & 1.39                               & 1.11                               & 1.24                               & 1.16                               & 1.04                               & 1.17                                \\ \cline{1-1}
 \multicolumn{1}{|l|}{\textbf{80\%}} & .99                                & 1.24                               & .85                                & .99                                & 1.47                               & 1.18                               & 1.27                               & 1.21                               & 1.14                               & 1.20                                \\ \cline{1-1}
 \multicolumn{1}{|l|}{\textbf{85\%}} & 1.18                               & 1.35                               & 1.00                               & 1.06                               & 1.65                               & 1.21                               & 1.43                               & 1.25                               & 1.22                               & 1.30                                \\ \cline{1-1}
 \multicolumn{1}{|l|}{\textbf{90\%}} & 1.22                               & 1.56                               & 1.18                               & 1.11                               & 1.81                               & 1.33                               & 1.66                               & 1.42                               & 1.31                               & 1.37                                \\ \cline{1-1}
 \end{tabular}
}
\caption{Percentile of Ratios in OML}
\label{tab:percentiles}
\vspace{-0.5cm}
\end{table}

\vspace{-0.25cm}


\section{Conclusion}

In this work, we presented the \name system. Based on a user's performance-centric SLA, \name scales the user's cluster of VMs in order to achieve good QoS at a low cost. We explored different techniques to layout the user's data to enable cluster resizing during a query session. We found that local, shared-nothing storage with partial data replication offers a practical solution with low setup times, minimal cluster resizing overheads, reliable query execution times, and low costs. \name further scales a user's cluster during a query session. While different scaling algorithms are possible, we find that perceptron learning yields results closest to those of an oracle and without difficult parameter tunings. As future work, we plan to combine reactive and proactive scaling techniques,  which could prove beneficial in cases where perceptron learning is not as accurate. 

\textbf{Acknowledgements} This project was supported in part by NSF grants IIS-1247469 and IIS-1524535, gifts from Amazon, the Intel Science and Technology Center for Big Data, and Facebook. J. Ortiz was supported in part by an NSF graduate fellowship.


\end{sloppypar}

\scriptsize
\bibliographystyle{abbrv}
\bibliography{header,elasticity}

\begin{thebibliography}{10}

\bibitem{amazonec2}
Amazon {EC2}.
\newblock \url{http://aws.amazon.com/ec2/}.

\bibitem{amazonaws}
Amazon {AWS}.
\newblock \url{http://aws.amazon.com/}.

\bibitem{spark}
Apache spark: Lightnight-fast cluster computing.
\newblock \url{http://spark.apache.org/}.

\bibitem{azure}
Microsoft {A}zure.
\newblock \url{http://azure.microsoft.com/en-us/}.

\bibitem{moa}
A.~Bifet et~al.
\newblock Moa: Massive online analysis, a framework for stream classification
  and clustering.
\newblock In {\em Journal of Machine Learning Research, 11:44–50}, 2010.

\bibitem{bigquery}
Google {B}ig{Q}uery.
\newblock \url{https://developers.google.com/bigquery/}.

\bibitem{Chi:11b}
Y.~Chi et~al.
\newblock Sla-tree: a framework for efficiently supporting sla-based decisions
  in cloud computing.
\newblock In {\em Proc. of the EDBT Conf.}, pages 129--140, 2011.

\bibitem{Chi:11a}
Y.~Chi, H.~J. Moon, and H.~Hacig{\"{u}}m{\"{u}}s.
\newblock icbs: Incremental costbased scheduling under piecewise linear slas.
\newblock {\em {PVLDB}}, 4(9):563--574, 2011.

\bibitem{Das:11}
S.~Das et~al.
\newblock Albatross: Lightweight elasticity in shared storage databases for the
  cloud using live data migration.
\newblock {\em {PVLDB}}, 4(8):494--505, 2011.

\bibitem{Elmore:14}
A.~J. Elmore, S.~Das, D.~Agrawal, and A.~{El Abbadi}.
\newblock Zephyr: live migration in shared nothing databases for elastic cloud
  platforms.
\newblock In {\em Proc. of the SIGMOD Conf.}, pages 301--312, 2011.

\bibitem{Elmore:13}
A.~J. Elmore, S.~Das, A.~Pucher, D.~Agrawal, A.~{El Abbadi}, and X.~Yan.
\newblock Characterizing tenant behavior for placement and crisis mitigation in
  multitenant dbmss.
\newblock In {\em Proceedings of the {ACM} {SIGMOD} International Conference on
  Management of Data, {SIGMOD} 2013, New York, NY, USA, June 22-27, 2013},
  pages 517--528, 2013.

\bibitem{Ferguson:12}
A.~D. Ferguson, P.~Bod{\'{\i}}k, S.~Kandula, E.~Boutin, and R.~Fonseca.
\newblock Jockey: guaranteed job latency in data parallel clusters.
\newblock In {\em EuroSys '12}.

\bibitem{Gandhi:16}
A.~Gandhi, P.~Dube, A.~Kochut, L.~Zhang, and S.~Thota.
\newblock Autoscaling for hadoop clusters.
\newblock In {\em IC2E 2016}.

\bibitem{Gupta:15}
A.~Gupta et~al.
\newblock Amazon redshift and the case for simpler data warehouses.
\newblock Proc. of the SIGMOD Conf., pages 1917--1923, 2015.

\bibitem{halperin:14}
D.~Halperin et~al.
\newblock Demonstration of the {M}yria big data management service.
\newblock In {\em SIGMOD}, pages 881--884, 2014.

\bibitem{Herodotou:11a}
H.~Herodotou et~al.
\newblock No one (cluster) size fits all: automatic cluster sizing for
  data-intensive analytics.
\newblock In {\em Proc. of the Second SoCC Conf.}, page~18, 2011.

\bibitem{Jalaparti:12}
V.~Jalaparti et~al.
\newblock Bridging the tenant-provider gap in cloud services.
\newblock In {\em Proc. of the 3rd ACM Symp. on Cloud Computing}, page~10,
  2012.

\bibitem{Janert:13}
P.~K. Janert.
\newblock {\em Feedback Control for Computer Systems}.
\newblock O'Reilly Media, Inc., 2013.

\bibitem{Kao:95}
B.~Kao et~al.
\newblock Advances in real-time systems.
\newblock chapter An Overview of Real-time Database Systems, pages 463--486.
  Prentice-Hall, Inc., 1995.

\bibitem{Karger:97}
D.~Karger et~al.
\newblock Consistent hashing and random trees: Distributed caching protocols
  for relieving hot spots on the world wide web.
\newblock STOC, pages 654--663, 1997.

\bibitem{Konstantinou:12}
I.~Konstantinou et~al.
\newblock {TIRAMOLA:} elastic nosql provisioning through a cloud management
  platform.
\newblock In {\em Proc. of the SIGMOD Conf.}, pages 725--728, 2012.

\bibitem{Kossmann:10}
D.~Kossmann, T.~Kraska, and S.~Loesing.
\newblock An evaluation of alternative architectures for transaction processing
  in the cloud.
\newblock In {\em Proceedings of the 2010 ACM SIGMOD International Conference
  on Management of Data}, SIGMOD '10, pages 579--590, New York, NY, USA, 2010.
  ACM.

\bibitem{Lim:10}
H.~Lim et~al.
\newblock Automated control for elastic storage.
\newblock In {\em {ICAC}}, pages 1--10, 2010.

\bibitem{Liu:03}
T.~Liu and M.~Martonosi.
\newblock Impala: A middleware system for managing autonomic, parallel sensor
  systems.
\newblock {\em SIGPLAN Not.}, 2003.

\bibitem{Liu:13}
Z.~Liu, H.~Hacig{\"{u}}m{\"{u}}s, H.~J. Moon, Y.~Chi, and W.-P. Hsiung.
\newblock Pmax: Tenant placement in multitenant databases for profit
  maximization.
\newblock In {\em Proceedings of the 16th International Conference on Extending
  Database Technology}, EDBT '13, New York, NY, USA, 2013. ACM.

\bibitem{Mahmoud:13}
H.~A. Mahmoud, H.~J. Moon, Y.~Chi, H.~Hacig{\"{u}}m{\"{u}}s, D.~Agrawal, and
  A.~{El Abbadi}.
\newblock Cloudoptimizer: multi-tenancy for i/o-bound {OLAP} workloads.
\newblock In {\em Joint 2013 {EDBT/ICDT} Conferences, {EDBT} '13 Proceedings,
  Genoa, Italy, March 18-22, 2013}, pages 77--88, 2013.

\bibitem{Minhas:12}
U.~F. Minhas et~al.
\newblock Elastic scale-out for partition-based database systems.
\newblock In {\em Proceedings of the 2012 IEEE 28th International Conference on
  Data Engineering Workshops}, ICDEW '12, pages 281--288, Washington, DC, USA,
  2012. IEEE Computer Society.

\bibitem{oneil:09}
P.~O'Neil, E.~O'Neil, and X.~Chen.
\newblock The star schema benchmark.
\newblock \url{http://www.cs.umb.edu/~poneil/StarSchemaB.PDF}.

\bibitem{ortiz:15}
J.~Ortiz et~al.
\newblock Changing the face of database cloud services with personalized
  service level agreements.
\newblock In {\em CIDR}, 2015.

\bibitem{Ortiz:16}
J.~Ortiz et~al.
\newblock Perfenforce demonstration: Data analytics with performance
  guarantees.
\newblock In {\em SIGMOD}, 2016.

\bibitem{Papaemmanouil:12}
O.~Papaemmanouil.
\newblock Supporting extensible performance slas for cloud databases.
\newblock In {\em Proc. of the 28th ICDE Conf.}, pages 123--126, 2012.

\bibitem{Rabl:10}
T.~Rabl, M.~Frank, H.~M. Sergieh, and H.~Kosch.
\newblock A data generator for cloud-scale benchmarking.
\newblock TPCTC'10, pages 41--56.

\bibitem{Ren:13}
K.~Ren, Y.~Kwon, M.~Balazinska, and B.~Howe.
\newblock Hadoop's adolescence: An analysis of hadoop usage in scientific
  workloads.
\newblock {\em Proc. VLDB Endow.}, 6(10):853--864, Aug. 2013.

\bibitem{Stonebraker:13}
M.~Stonebraker and A.~Weisberg.
\newblock The voltdb main memory {DBMS}.
\newblock {\em {IEEE} Data Eng. Bull.}, 36, 2013.

\bibitem{Sutton:98}
R.~S. Sutton and A.~G. Barto.
\newblock Reinforcement learning i: Introduction, 1998.

\bibitem{Venkataraman:16}
S.~Venkataraman, Z.~Yang, M.~J. Franklin, B.~Recht, and I.~Stoica.
\newblock Autoscaling for hadoop clusters.
\newblock In {\em NSDI 2016}.

\bibitem{Vo:10}
H.~T. Vo, C.~Chen, and B.~C. Ooi.
\newblock Towards elastic transactional cloud storage with range query support.
\newblock {\em {PVLDB}}, 3(1):506--517, 2010.

\bibitem{Xiong:11a}
P.~Xiong et~al.
\newblock Activesla: a profit-oriented admission control framework for
  database-as-a-service providers.
\newblock In {\em Proc. of the Second SoCC Conf.}, page~15, 2011.

\end{thebibliography}

\end{document}